\documentclass[Review,sagev,times]{sagej}

\usepackage{moreverb,url}
\usepackage{epstopdf}

\usepackage[colorlinks,bookmarksopen,bookmarksnumbered,citecolor=red,urlcolor=red]{hyperref}

\newcommand{\rmd}{{\rm d}}
\newcommand{\rme}{{\rm e}}
\def\volumeyear{2021}
\begin{document}

\runninghead{Davies et al}

\title{Retarded kernels for longitudinal survival analysis and dynamic prediction}

\author{Annabel L Davies\affilnum{1}, Anthony C C Coolen\affilnum{2}\affilnum{3} and Tobias Galla\affilnum{1}\affilnum{4}}

\affiliation{\affilnum{1}Department of Physics and Astronomy, University of Manchester, UK\\
\affilnum{2}Department of Biophysics, Radboud University, The Netherlands\\
\affilnum{3}Saddle Point Science Ltd, UK\\
\affilnum{4}Instituto de F\'isica Interdisciplinar y Sistemas Complejos, IFISC (CSIC-UIB), Campus Universitat Illes Balears, Palma de Mallorca, Spain}

\corrauth{Annabel L Davies, Theoretical Physics, Department of Physics and Astronomy, School of Natural Sciences, The University of Manchester, Manchester M13 9PL, United Kingdom}

\email{annabel.davies@postgrad.manchester.ac.uk}

\begin{abstract}
{Predicting patient survival probabilities based on observed covariates is an important assessment in clinical practice. These patient-specific covariates are often measured over multiple follow-up appointments. It is then of interest to predict survival based on the history of these longitudinal measurements, and to update predictions as more observations become available. The standard approaches to these so-called `dynamic prediction' assessments are joint models and landmark analysis. Joint models involve high-dimensional parameterisations, and their  computational complexity often prohibits including multiple longitudinal covariates. Landmark analysis is simpler, but discards a proportion of the available data at each `landmark time'.  In this work we propose a `retarded kernel' approach to dynamic prediction that sits somewhere in between the two standard methods in terms of complexity. By conditioning hazard rates directly on the covariate measurements over the observation time frame, we define a model that takes into account the full history of covariate measurements but is more practical and parsimonious than joint modelling. Time-dependent association kernels describe the impact of covariate changes at earlier times on the patient's hazard rate at later times. Under the constraints that our model (i) reduces to the standard Cox model for time-independent covariates, and (ii) contains the instantaneous Cox model as a special case, we derive two natural kernel parameterisations. Upon application to three clinical data sets, we find that the predictive accuracy of the retarded kernel approach is comparable to that of the two existing standard methods.  }
\end{abstract}


\maketitle

\section{Introduction}

Survival analysis is a well established field of medical statistics that involves modelling the probability of survival until some specified irreversible event such as death or the onset of disease. Of particular clinical interest is the prediction of patient-specific survival based on a set of observed biomarkers or `covariates'.\cite{vanH:2007}  Such predictions aid clinicians in making treatment and testing decisions, and provide personalised  information for patients about their health.\cite{Rizopoulos:2017}  

Cox's proportional hazards (PH) model\cite{Cox:1972} remains the most widely used model in survival analysis.\cite{Tian:2005, Lee:2019} In this context, survival is assumed to depend on a set of covariates, $z_\mu$, $\mu=1,\hdots,p$, measured at some baseline time. The hazard, $h(t)$, is the probability per unit time of the event happening at time $t$ given that no event has occurred up to that time. In Cox's PH model this hazard is defined as
\begin{align}
\label{eq:Cox}
    h(t) = h_0(t)  \rme^{\sum_{\mu=1}^{p} \beta_\mu z_\mu},
\end{align}
where the $\beta_\mu$ (with $\mu=1,\hdots, p$) are the so-called association parameters. The base hazard rate, $h_0(t)$, is the value of the hazard for covariate values $z_\mu=0$ $\forall \mu$. The name ‘proportional hazards’ refers to the fact that, due to the exponential form of the hazard function, the effect of each covariate on the hazard is multiplicative. In this work we will call the model in Equation~(\ref{eq:Cox}) the `standard Cox model'. 

Survival prediction in the standard Cox model is based on the survival function, 
\begin{align}
    S(t) =  \rme^{-\int_0^t \rmd t' h(t')},
\end{align}
that describes the probability that an individual with hazard function $h(\cdot)$ experiences the event after time $t$.

In reality, covariates are often measured repeatedly over time. This means that multiple observations of time-dependent covariates $\{z_\mu(t)\}$  are made for any  particular patient. A simple extension to the standard Cox model involves modelling the hazard rate as dependent on the instantaneous value of the covariates,\cite{Kleinbaum:2005, Kalbfleisch:2012, Mills:2012} that is
\begin{align}
\label{eq:Instant}
    h(t) = h_0(t)  \rme^{\sum_{\mu=1}^p \beta_\mu z_\mu(t) }.
\end{align}
We refer to this as the `instantaneous Cox model'. 

However, in practice, one does not have access to the full covariate trajectories $z_\mu(t)$. Instead observations are made at discrete follow-up times until some subject-specific final observation time. Since we do not have access to covariate measurements after this time, we cannot make predictions about future survival probabilities based on Equation~(\ref{eq:Instant}). Of particular difficulty is the inclusion of so-called `endogenous' covariates.\cite{Rizo:2012}

Due to these difficulties, survival predictions are commonly evaluated by treating the baseline covariate measurements as fixed values in a standard Cox model.\cite{Rizopoulos:2017} By not including the follow-up observations, this standard practice discards a potentially considerable proportion of the available patient data.  

Recently, there has been much interest in so-called `dynamic prediction'.\cite{vanH:2007, Rizopoulos:2011, vanH:2011} These methods aim to make survival predictions based on the longitudinal history of biomarker data, and update these predictions as more data becomes available. Such analysis is clinically valuable as it allows patients and clinicians to review disease progression over time and update the prognosis at each follow-up visit.\cite{Zhu:2018} Currently, there are two main approaches to dynamic prediction; joint modelling and landmarking. 

Landmarking was an early approach to the problem,\cite{Anderson:1983} whereby a standard Cox model is fitted to patients in the original data set who are still at risk at the time point of interest, using their most recent covariate measurements. 

More recently, joint modelling has become an established method.\cite{Tsiatis:1995, Ibrahim:2010, Wu:2012, Rizo:2012} Here one models the time-dependent covariate trajectory using a parameterised longitudinal model, and this complete trajectory is then inserted into an instantaneous Cox-type survival model. A joint likelihood of the longitudinal and survival sub-models is constructed, and the model parameters are estimated via maximum likelihood or Bayesian inference. 

Both methods have limitations. In particular, joint models are demanding both conceptually and computationally. Correctly modelling the longitudinal trajectories can be difficult when patient measurements exhibit varied non-linear behaviour\cite{Zhu:2018} and misspecification of this trajectory has been found to lead to bias.\cite{Arisido:2019} Furthermore, the number of model parameters increases rapidly with the inclusion of multiple longitudinal markers. This means that many software packages cannot handle more than one longitudinal covariate,\cite{Moreno-Betancur:2018, Gould:2015, Hickey:2016} and those that can quickly become computationally intensive.\cite{Mauff:2020, Li:2017} For these reasons, the landmarking model is often seen as the only practical option.\cite{Zhu:2018} However, the relative simplicity of the landmarking approach comes with its own drawbacks. By using only the `at risk' data set to make predictions at a certain time (discarding patients who had an event before the landmark time), landmarking makes use of only a subset of the available data. In standard landmarking approaches, the history of the covariate values are not taken into account directly, and a new model must be fitted every time one wishes to update the predictions.

In this work we present a new approach to dynamic prediction that conceptually and in terms of computational complexity lies somewhere in between the joint modelling  and landmarking methods. Rather than modelling the covariate trajectory at future times, as in the joint modelling approach, we model the probability of survival conditioned directly on the observed covariates measured from the baseline time up to a subject-specific final observation time. Unlike the landmark approach, a single model is fitted to all of the available data, using the full history of the covariate values. We do, however, maintain well-established and desirable features of the Cox model, so that our model contains the instantaneous Cox model as a special case, and reduces automatically to the standard Cox model for covariates that are observed to be fixed over time. Within these constraints, we define time-dependent parametric association kernels, $\beta_\mu(t,t',s)$, that describe the impact of changes of covariate $\mu$ at time $t^\prime$ on patient risk at some later time $t$. The kernel can also depend on the final observation time $s$ for the patient. Building on ideas from weighted cumulative exposure models,\cite{Breslow:1983, Thomas:1988} these kernels allow us to assign smaller effects to covariates that were measured further in the past.  We refer to our method as the `retarded kernel' approach.

The remainder of this article is set out as follows. In Section~\ref{Data} we introduce the motivating data sets.  In Section~\ref{Models} we then provide details of the dynamic prediction models. We begin by describing the longitudinal and time-to-event data, and briefly outline the standard methods: joint modelling (Section~\ref{JointModels}) and landmarking (Section~\ref{landmark}). In Section~\ref{OurModels}, we introduce the retarded kernel approach. We start by defining the hazard rate conditioned on the observed data, and then develop two natural parameterisations for the association kernels that meet our requirements. We outline the maximum likelihood method for parameter estimation for these models, and show how the retarded kernel approach can be used to make dynamic predictions. Via application to the real data sets, in  Section~\ref{Apply} we compare the performance of the retarded kernel approach to the standard methods using an established measure of predictive accuracy. Finally, we discuss and summarise our results in Section~\ref{Discuss}.

\section{Motivating data sets}
\label{Data}
In our work we will assess the predictive capabilities of the different models for dynamic prediction using three clinical data sets, that contain both longitudinal covariate measurements and time-to-event data. All three data sets are publicly available in the \texttt{JMbayes} package,\cite{Rizo:2016} and were used in Rizopoulos (2012)\cite{Rizo:2012} to illustrate the joint modelling  method.

\subsection{Primary biliary cirrhosis}
\label{PBC}
\begin{figure}[b]
    \centering
    \includegraphics[width=1\linewidth]{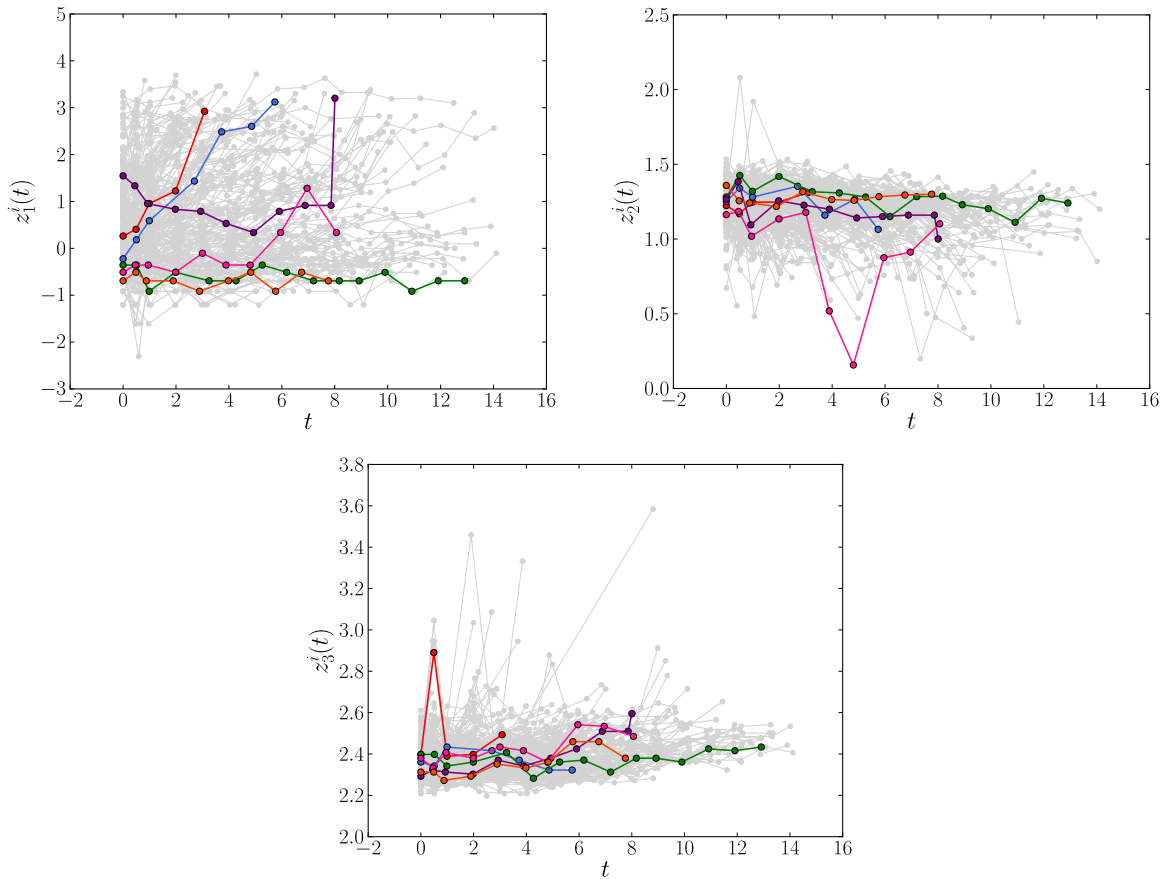}
    \caption{~ The longitudinal profiles of the time-dependent covariates log serum bilirubin ($z^i_1(t)$), log serum albumin ($z^i_2(t)$) and log prothrombin time  ($z^i_3(t)$) for the $N=312$ patients ($i=1,\hdots, N$) in the PBC data set described in Section~\ref{PBC}. For clarity, the trajectories of 6 individuals are highlighted. Time, $t$, on the $x$-axis is measured in years. }
    \label{fig:Mayo}
\end{figure}

The first motivating data set is a from a study conducted by the Mayo Clinic from 1974 to 1984 on patients with primary biliary cirrhosis (PBC), a progressive chronic liver disease.\cite{Murtaugh:1994} We will refer to this as the PBC data. The study involved $N=312$ patients who were randomly assigned either a placebo (154 patients) or the D-penicillamine treatment (158 patients). Time-to-event data is available for the outcome of interest (death) or the censoring event (either the time at which the patient receives a liver transplant or the final follow-up time at which they were still alive). By the end of follow-up, 140 patients had died, 29 had received a transplant and 143 were still alive. As well as baseline covariate measurements such as age at baseline and gender, multiple longitudinal biomarker measurements were collected for each patient over an average number of 6.2 visits from study entry to some subject-specific final observation time (prior to their event time). While the original aim of the study was to investigate the effect of the drug D-penicillamine, no effect was found and the data has since been used to study the progression of the disease based on longitudinal biomarkers.\cite{Schoop:2008} With this in mind, we include age at baseline as our only fixed covariate, and focus on the longitudinal covariates log serum bilirubin, log serum albumin and log prothrombin time, which have previously been found to be indicators of patient survival.\cite{Schoop:2008} Serum bilirubin and serum albumin indicate concentrations of these substances in the blood, measured in mg/dl and g/dl respectively. Prothrombin time measures the time (in seconds) it takes for blood to clot in a sample. Time series of these three longitudinal biomarkers are plotted in Figure~\ref{fig:Mayo}.

\subsection{AIDS}
\label{AIDS}

\begin{figure}[h]
    \centering
    \includegraphics[width=0.5\linewidth]{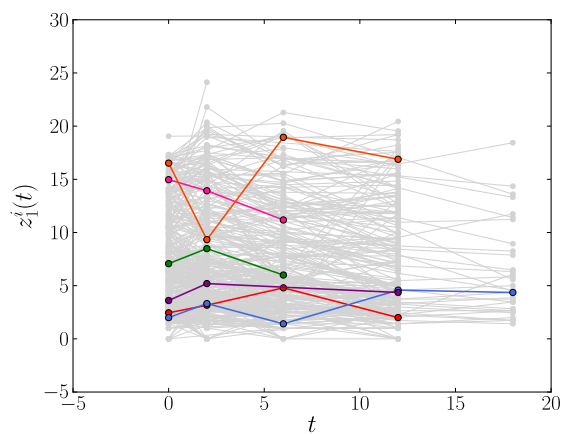}
    \caption{~ The longitudinal profiles of CD4 count ($z^i_1(t)$) in the $N=467$ patients ($i=1,\hdots, N$) in the AIDS data set in Section~\ref{AIDS}. For clarity, the trajectories of 6 individuals are highlighted. Time, $t$ is measured in months.  }
    \label{fig:CD4}
\end{figure}

The second data set involves $N=467$ HIV-infected patients who had failed to respond, or were intolerant to,  zidovudine (previously called `azidothymidine')  therapy (AZT).\cite{Abrams:1994} The aim of the study was to compare two antiretroviral drugs, didanosine (ddI) and zalcitabine (ddC). Patients were randomly assigned one of these drugs at baseline. Patients' CD4 cell counts were recorded at baseline and follow-up measurements were planned at 2, 6, 12 and 18 months. CD4 cells are white blood cells that  fight infections. A decrease in the number of CD4 cells over time indicates a worsening of the immune system and higher susceptibility to infection. Therefore,  the number of CD4 cells in a blood sample is an important marker of immune strength and hence a covariate of interest in HIV-infected patients. By the end of the study 118 patients had died, and the time to event (death) or censoring was recorded for all patients. Final observation times ($s_i\in[0, 2, 6, 12, 18]$ months) were always less than their corresponding event times, such that there is a time gap between when a subject was last observed and when they experienced an event. Following Guo and Carlin (2004),\cite{Guo:2004} we included, in addition to the longitudinal CD4 counts and the patients' drug group, also three other binary fixed covariates in our analysis: gender, PrevOI (previous opportunistic infection -- AIDS diagnosis -- at study entry), and Stratum (whether the patient experienced AZT failure or AZT intolerance). We will refer to this data as the AIDS data set. The longitudinal profiles of the CD4 count for all patients are plotted in Figure~\ref{fig:CD4}.

\subsection{Liver cirrhosis}
\label{Liver}

\begin{figure}[h]
    \centering
    \includegraphics[width=0.5\linewidth]{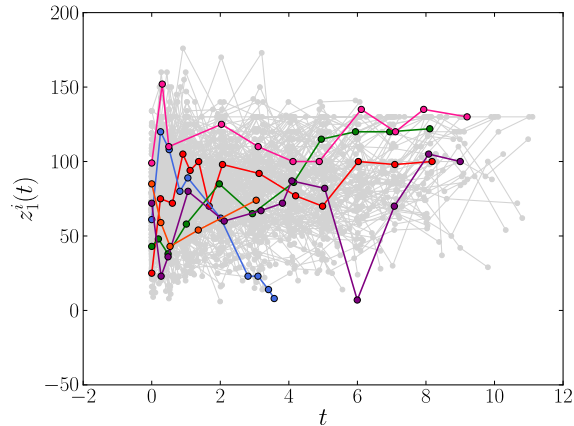}
    \caption{~ The longitudinal profiles of the prothrombin index ($z^i_1(t)$) as measured in the $N=488$ patients ($i=1,\hdots,N)$ in the Liver data set described in Section~\ref{Liver}. For clarity, the trajectories of 6 individuals are highlighted. Time, $t$, is measured in years.  }
    \label{fig:Proth}
\end{figure}

The third data set is from a trial conducted between 1962 and 1974, involving $N=488$ patients with liver cirrhosis, a general term including all forms of chronic diffuse liver disease.\cite{Andersen:1993} We call this the Liver data set. At baseline, 251 patients were randomly assigned a placebo and 237 were assigned treatment with the drug prednisone. Follow-up appointments were scheduled at 3, 6 and 12 months and then yearly thereafter, though actual follow up times varied considerably. At these follow up appointments, multiple longitudinal biomarkers were measured. However, only the prothrombin index measurements are available from the \texttt{JMbayes} package. This is a measure of liver function based on a blood test of coagulation factors produced by the liver. For reproducibility, and following previous analyses of the Liver data set,\cite{Henderson:2002, Rizo:2012} we include the prothrombin index as our only time-dependent biomarker. The drug group is included as a fixed baseline covariate. By the end of the study, 150 prednisone-treated, and 142 placebo-treated patients had died. Their time-to-event data was recorded. Of the 488 subjects, 120 were observed until their event time while all others were observed until some subject-specific final observation time before their event time. Figure~\ref{fig:Proth} shows the longitudinal prothrombin measurements for all patients. 

\section{Dynamic prediction models}
\label{Models}
\subsection{Setup and notation}
\label{SetUp}
In this work, we consider longitudinal survival data  of the form $\mathcal{D} = \{T_i, \delta_i, \mathcal{Z}^i; i=1,\hdots,N \}$ where $T_i=\min(T_i^*,C_i)$ is the observed event time of individual $i$ with $T_i^*$ denoting the true event time and $C_i$ denoting the censoring time. The event indicator $\delta_i=I(T_i^*\leq C_i)$ is equal to 1 when the true event time is observed and 0 when it is censored. Throughout this article we use the indicator function $I(A)$, defined as $I(A)=1$ if $A$ holds, and  $I(A)=0$ otherwise. $\mathcal{Z}^i=\{z^i_\mu(t_{i\ell}); ~\mu=1,\hdots,p, ~\ell=1,\hdots,n_i,  ~t_{i\ell} \in [0,s_i]\}$  denotes the set of time-dependent covariate observations of individual $i$. Individual $i$ has $n_i$ measurements of $p$ longitudinal covariates from time $t=0$ up to some subject-specific final observation time $s_i\leq T_i$. These measurements are taken at discrete (subject-specific) observation times, $t_{i\ell}$, $\ell=1,\hdots,n_i$, where $t_{i1}=0$ and $t_{i n_i}=s_i$. We write $\mathcal{Z}^i_{[0,s_i]} = \{z^i_\mu(t);~\mu=1\ldots p,  ~t\in [0,s_i]\}$ for the `true' (but non-accessible) continuous trajectories of the $p$ covariates over the interval $t \in [0,s_i]$ for individual $i$. We develop our theory based on the assumption that we have access to these trajectories $\mathcal{Z}^i_{[0,s_i]}$. As  we will see later, we estimate $\mathcal{Z}^i_{[0,s_i]}$ from the discrete observations $\mathcal{Z}^i$.

We are interested in predicting survival probabilities for some new subject with longitudinal measurements $\mathcal{Z}=\{z_\mu(t_{\ell}); ~\mu=1,\hdots,p, ~\ell=1,\hdots,n,  ~t_{\ell} \in [0,s]\}$. The quantity we wish to estimate is the probability that this subject survives until some future time $u>s$, conditional on their survival to $s$, and on their covariate observations up to $s$. That is,
\begin{align}
\label{eq:Pi-1}
    \pi(u|\mathcal{Z}_{[0,s]},s) = \mathrm{Pr}(T^*\!\geq u | ~T^*\! > s, \mathcal{Z}_{[0,s]}, \mathcal{D}).
\end{align}
The quantity $\pi(u|\mathcal{Z}_{[0,s]},s)$ is referred to as a `dynamic predictor' due to the fact that it can be updated as more measurements become available at later times.\cite{vanH:2011, Rizopoulos:2017}

\subsection{Joint Models}
\label{JointModels}

In joint modelling one specifies two model components: a longitudinal model for the trajectory of the time-dependent covariates, and a survival model which relates to the covariate trajectory via shared parameters. In the \texttt{JMbayes} package, joint models are fitted using Bayesian inference by specifying a joint likelihood distribution for the two model components and a set of prior distributions on the model parameters.  Details of this package and the joint modelling framework we follow are described in Rizopoulos (2016)\cite{Rizo:2016} and Rizopoulos (2012).\cite{Rizo:2012} In this section we briefly outline the model.

\subsubsection{Longitudinal modelling component.}
Mixed-effects models are typically specified for the longitudinal covariate trajectories, where it is assumed that the observed value $z_\mu (t)$ of the covariate at time $t$ deviates from the true (unobserved) value $m_\mu(t)$ by an amount $\varepsilon_\mu(t)$. 
The error terms $\varepsilon_\mu(t)$ of all subjects are assumed to be statistically independent, and normally distributed with variance $\sigma_\mu^2$:
\begin{align}
    &z^i_\mu (t) = m^i_\mu (t) + \varepsilon^i_\mu (t),~~~~
   m^i_\mu (t)  = \boldsymbol{x}^{i\top}_\mu(t) \boldsymbol{\eta}_\mu + \boldsymbol{y}^{i\top}_\mu (t) \boldsymbol{b}^i_\mu\\
    &\boldsymbol{b}^i \sim \mathcal{N}(\boldsymbol{0},\boldsymbol{D}), \hspace{10 pt} \varepsilon^i_\mu (t) \sim \mathcal{N}(0,\sigma_\mu^2).
    \nonumber 
\end{align}
Between-subject variability is modelled via estimation of subject-specific random effects $\boldsymbol{b}^i_\mu$, whereas effects that are shared between all subjects are modelled by the fixed effects $\boldsymbol{\eta}_\mu$. The vectors $\boldsymbol{x}^{i\top}_\mu(t)$ and $\boldsymbol{y}^{i\top}_\mu (t)$ denote the design vectors for these fixed and random effects respectively. For multivariate models, one can allow for association between the different longitudinal markers via their corresponding random effects. In particular, we assume that the complete vector of random effects $\boldsymbol{b}^i=(\boldsymbol{b}^{i\top}_1,\hdots,\boldsymbol{b}^{i\top}_{p})^\top$ follows a multivariate normal distribution with mean zero and  variance-covariance matrix $\boldsymbol{D}$ that describes the correlations between and variances of the random effects. For details we refer to References  \cite{Rizopoulos:2017, Rizopoulos:2011, Rizo:2012, Cekic:2021, Mauff:2020}.

\subsubsection{Survival modelling component.}
The hazard at time $t$ is assumed to depend on the value of the longitudinal covariate at time $t$ without measurement error, that is
\begin{align}
\label{eq:JM_haz}
    h^{\mathrm{JM}}(t|\mathcal{M}_{[0,t]}) = h_0(t) \exp \left\{  \sum_{\mu=1}^{p} \alpha_\mu m_\mu (t) \right\},
\end{align}
where $\mathcal{M}_{[0,t]}=\{m_\mu(t'); ~\mu=1,\hdots,p, ~t'\in[0,t]\}$ denotes the history of the `true' (unobserved) longitudinal covariates up to time $t$.  Note that in Equation~(\ref{eq:JM_haz}) the hazard rate depends only on the instantaneous values of the covariates, but this can be generalised as briefly highlighted below. Unlike in the Cox model, the baseline hazard, $h_0(t)$ cannot be expressed analytically in terms of the other model parameters during the maximum likelihood procedure, but must instead be specified. Often this is done using a flexible parametric model, for example using penalised spline functions.\cite{Rizo:2016} Dependence of the hazard function on time-independent covariates $\{\zeta_\nu; ~\nu=1,\hdots,q\}$ can be included through an additional term $\sum_{\nu=1}^{q} \gamma_\nu \zeta_\nu$ in the exponent in Equation~(\ref{eq:JM_haz}), where $\gamma_\nu$ is the association parameter for fixed covariate $\nu$. Alternative extensions allow the hazard to depend on the slope of the covariate trajectory, or on its cumulative effect, by replacing  the term $\alpha_\mu m_\mu(t)$ with $\alpha_{\mu}^{(1)} m_\mu(t) + \alpha_{\mu}^{(2)} {\frac{\rmd}{\rmd t}m_\mu}(t)$ or with $\alpha_\mu \int_0^t\!\rmd t^\prime~ m_\mu(t^\prime)$,  respectively.\cite{Rizo:2012, Rizopoulos:2017} 

It has also been proposed to introduce a weight function to capture cumulative effects, writing $\alpha_\mu \int_0^t\!\rmd t^\prime ~ w_\mu(t-t^\prime) m_\mu(t^\prime)$, with kernels $w_\mu(t-t^\prime)$ defined such that earlier covariate values have a smaller effect on the hazard than recent values.\cite{Rizo:2012, Mauff:2017} This idea is connected to the concept of `weighted cumulative exposure' (WCE).\cite{Breslow:1983, Thomas:1988} WCE models were developed in etiological research to describe the complex cumulative effect of time-dependent `exposure', e.g. to a drug, on health outcomes.\cite{Abrahamowicz:2012} In a survival context, these models rely on continuous knowledge of the exposure all the way up to the event time,  and have hence been used almost exclusively for measuring the effects of external exposures such as treatments or environmental factors.\cite{Sylvestre:2009} Joint modelling allows the principles of WCE to be used for any longitudinal covariate. Through the prediction of future covariate trajectories, it is also possible for these ideas to be integrated into dynamic predictions.\cite{Mauff:2017} As we will explain later, we build on the principles of WCE to develop our retarded kernel approach.

\subsubsection{Dynamic prediction.}
\label{JM_Pred}
Finally, in the joint modelling framework we use, the quantity $\pi(u|\mathcal{Z}_{[0,s]},s)$ is estimated using a Bayesian approach, with posterior parameter distribution $p(\boldsymbol{\theta}_{\mathrm{JM}}|{\mathcal D})$ and where $\boldsymbol{\theta}_{\mathrm{JM}}$ is the vector of all the model parameters in the joint model. This leads to the estimator
\begin{align}
\label{eq:Pi-JM}
    \hat{\pi}^{\mathrm{JM}}(u|\mathcal{Z}_{[0,s]},s) = \int \mathrm{Pr}(T^*\!\geq u| ~T^*\!>s, \mathcal{Z}_{[0,s]}, \boldsymbol{\theta}_{\mathrm{JM}}) p(\boldsymbol{\theta}_{\mathrm{JM}}|\mathcal{D}) ~\mathrm{d} \boldsymbol{\theta}_{\mathrm{JM}}.
\end{align}
The parameter average in Equation~(\ref{eq:Pi-JM}) can generally not be evaluated analytically, and is computed via Monte Carlo methods. Again, we refer to References\cite{Rizopoulos:2017, Rizo:2012, Rizopoulos:2011} for details.

\subsubsection{Limitations.}
Joint models have undergone much development over recent years, with various extensions making the approach flexible in a range of different scenarios. However, the joint modelling approach requires the ability to correctly specify both the longitudinal and survival model. This can involve modelling assumptions which are not always easy to verify. Indeed, simulations have demonstrated that the joint modelling approach is biased under misspecification of the longitudinal model.\cite{Arisido:2019}
In addition, as more longitudinal outcomes are included, the dimensionality of the random effects increases, and fitting the joint model becomes computationally intensive. Depending on the longitudinal model specified, it can be difficult to include more than three or four longitudinal covariates.\cite{Mauff:2020, Zhu:2018} This is amplified when cumulative or weighted cumulative effects are used in the survival model (as numerical integration of the longitudinal model is required). As a result, there are cases where joint models are not a viable option and, instead, one must rely on approaches such as landmarking.\cite{Zhu:2018}

\subsection{Landmarking}
\label{landmark}
\subsubsection{Description of the landmarking procedure.}
The landmarking approach to dynamic prediction is based on the standard Cox model.\cite{Anderson:1983, vanH:2007, vanH:2011} Upon denoting with $\mathcal{R}(\upsilon) = \{i: ~T_i>\upsilon\}$ the set of individuals in the original data set who are still at risk at time $\upsilon$,  the landmarking model assumes that for a subject in the risk set $\mathcal{R}(\upsilon)$ the distribution of survival times, conditioned on the covariate measurements $\{z^i_\mu(\upsilon)\}$ at that time, follows a standard Cox model.\cite{Zhu:2018} In general one does not have covariate measurements for all individuals at time $\upsilon$. Instead, one uses for each individual the last observation $\{\tilde{z}_\mu(\upsilon),~\mu=1,\hdots, p\}$ of the covariates before time $\upsilon$, and treats these as fixed covariates in a standard Cox model with $\upsilon$ as the baseline time. That is, for the so-called `landmark time' $\upsilon$ one defines the hazard rate at times $t>\upsilon$ as
\begin{align}
\label{eq:LM_model}
    h^{\mathrm{LM}}(t | \mathcal{Z},\upsilon) = h_0 (t|\upsilon) \exp\left\{  \sum_{\mu=1}^{p} \alpha_\mu(\upsilon) \tilde{z}_\mu(\upsilon) \right\}.
\end{align}
The baseline hazard function $h_0(t|\upsilon)$ is unspecified, and is estimated as in standard Cox models, via partial likelihood arguments, or via functional maximisation of the data log-likelihood. Subsequently the association parameters are estimated. This procedure is carried out for each choice of the landmark time $\upsilon$, and leads to the Breslow estimator\cite{Breslow:1972} $\hat{h}_0(t|\upsilon)$ and the association parameters $\hat\alpha_\mu(\upsilon)$. The main difference compared to standard Cox models is the dependence of association parameters and the base hazard rate on the landmark time $\upsilon$.

To estimate the quantity $\pi(u|\mathcal{Z}_{[0,s]},s)$, the landmark time $\upsilon$ in  Equation~(\ref{eq:LM_model}) is set equal to $s$. Once this model is fitted, survival prediction to time $u>s$ is performed using the standard Cox survival probability,
\begin{align}
\label{eq:LM_pi}
    \hat{\pi}^{\mathrm{LM}}(u|\mathcal{Z}_{[0,s]}, s) = \exp\Bigg\{ -  \rme^{\sum_{\mu=1}^{p} \hat\alpha_\mu(s) \tilde{z}_\mu(s)}\int_{s}^u \hat{h}_0(t'|s) \rmd t^\prime \Bigg\}.
\end{align}

\subsubsection{Limitations.}
Landmarking is computationally and conceptually much simpler than the joint modelling approach. For data sets with multiple longitudinal covariates, disparate non-linear covariate trajectories or categorical time-dependent covariates, landmarking is often the preferred approach.\cite{Zhu:2018} However, it also has limitations. For example, the model focuses only on the most recent value observed before time $\upsilon$, and does not account for the earlier history of covariates. Furthermore, data from individuals who experience the event before time $\upsilon$ is not used for the parameter estimation at landmark time $\upsilon$. Therefore, the landmark approach uses only a subset of the available data. In addition, a new Cox model has to be specified and fitted for each landmark time. Therefore, in order to update predictions after each time where subject $j$ is observed, one must refit the model using a new risk set. The longer subject $j$ is observed, the fewer individuals remain in the risk set and less data is available to do this.

\subsection{Retarded kernel approach}
\label{OurModels}

We now introduce our retarded kernel approach to dynamic prediction. It aims to overcome some of the limitations of the standard joint modelling and landmarking methods. Unlike landmarking, the retarded kernel approach aims to incorporate the entire data set, including the full history of covariate values while, at the same time remaining conceptually and computationally simpler than joint models.

\subsubsection{General setup.}
The starting point for the retarded kernel approach is an expression for the hazard rate that resembles that of weighted cumulative exposure models,\cite{Breslow:1983, Thomas:1988} 
\begin{equation}\label{eq:retkernel}
h^{\mathrm{RK}}(t|{\cal Z}_{[0,s]})=h_0(t)\exp\left\{\int_0^{\mbox{\footnotesize{min}}(s,t)}\sum_{\mu=1}^{p} \beta_\mu(t,t^\prime\!,s)z_\mu(t^\prime)\rmd t^\prime \right\}.
\end{equation}
In this expression the $\{z_\mu(t')\}$ are time-dependent covariates, which we assume to be known from time $0$ up to time $s$. To keep the notation compact we have left out time-independent covariates, as these can always be included trivially. This model differs from the joint model approach to WCE in how we deal with covariates that are only observed up to some final observation time $s$ before the event time.  When $t\leq s$ (i.e. when  $t$ is a point in time prior to the last observation of covariates) the hazard rate in Equation~(\ref{eq:retkernel}) only depends on covariates up to time $t$. For times $t\geq s$ covariates up to time $s$ enter into the hazard rate.

The kernel $\beta_\mu(t,t^\prime\!,s)$ describes (potentially) retarded effects of covariates. More precisely, $\beta_\mu(t,t^\prime\!,s)$ quantifies the effect of the value of covariate $\mu$ at time $t^\prime$ on the hazard rate at a later time $t$, for a patient whose covariates are known up to time $s$. The form of Equation~(\ref{eq:retkernel}) ensures causality, since only covariate values at times $t^\prime\leq t$  contribute to the hazard at time $t$. We set $\beta(t,t',s)=0$ for $t'>t$. In principle, the precise form of $\beta_\mu(t,t^\prime\!,s)$  could be chosen from a wide range of functions. We reduce this freedom  
via the following requirements which must hold for all $\mu$:
\begin{enumerate}
\item[(i)] {\em Exponential decay of covariate impact.} 
 We assume that the impact of each covariate $\mu$ at time $t^\prime$ on the hazard rate at a later time $t>t^\prime$ decays exponentially with the time difference $t-t^\prime$. How fast the effect of the covariate decays is governed by a covariate-specific impact time scale $\tau_\mu\geq 0$. 
 \item[(ii)] {\em Equivalence with standard Cox model for stationary covariates.} 
Our second requirement is 
that expression (\ref{eq:retkernel}) reduces to the standard Cox model in Equation~(\ref{eq:Cox}) in the case of a constant covariate, i.e. when $z_\mu(t)\equiv z_\mu$ for all $t$. This is achieved when there is a constant $a_\mu$, which is independent of $t$ and $s$, such that 
\begin{equation}\label{eq:constcond}
   \int_0^{\mbox{\footnotesize{min}}(s,t)}\! \beta_\mu(t,t^\prime\!,s)\rmd t^\prime=a_\mu.
\end{equation}
\item[(iii)] {\em Equivalence with instantaneous Cox model for short impact time scales.}
Finally,  for $0<t\leq s$ we require that expression (\ref{eq:retkernel}) reduces to the instantaneous Cox model in Equation~(\ref{eq:Instant}) in the limit $\tau_\mu\downarrow 0$, i.e. when the covariate impact on risk decays immediately. This is achieved, without violating (ii), if we have 
\begin{equation}
\lim_{\tau_\mu\downarrow 0} \beta_\mu(t,t^\prime\!,s)=a_\mu\delta(t-t^\prime).
\end{equation}
\end{enumerate}

From (i) it follows that our kernel $\beta_\mu(t,t^\prime,s)$ must have the following form:
\begin{eqnarray}
\beta_\mu(t,t^\prime,s)= A_\mu(t,s)\tau_\mu^{-1}\rme^{-(t-t^\prime)/\tau_\mu}+B_\mu(t,s),
\label{eq:general_form}
\end{eqnarray}
where the quantities $A_\mu(t,s)$ and $B_\mu(t,s)$ can depend on $\tau_\mu$ in general. Requirements (ii) and (iii) then translate into, respectively, 
\begin{eqnarray}
s,t\geq 0: &~~&  A_{\mu}(t,s)\rme^{-t/\tau_\mu}\Big(\rme^{{\rm min}(s,t)/\tau_\mu}-1\Big)+{\rm min}(s,t)B_{\mu}(t,s)=a_\mu
\label{eq:condition1}
\\
0<t\leq s:  &~~& \lim _{\tau_\mu\downarrow 0}A_\mu(t,s)=a_\mu,~~\lim _{\tau_\mu\downarrow 0}B_\mu(t,s)=0.
\label{eq:condition2}
\end{eqnarray}

\subsubsection{Two models within this family.}\label{sec:simple} 
Finally, from the remaining family of models, i.e. those that satisfy Equations (\ref{eq:condition1}) and (\ref{eq:condition2}), we choose the two simplest members. These are defined by demanding that either $B_\mu(t, s)=0$ for {\em any} $\tau_\mu$ (model A), or that $A_\mu(t,s)=a_\mu$ for {\em any} $\tau_\mu$ (model B). Working out the details for these choices via Equations (\ref{eq:condition1}, \ref{eq:condition2}) then leads to the following formulae:
\begin{eqnarray}
\mbox{Model A:}&~~& \beta^A_\mu(t,t^\prime\!,s)=\frac{a_\mu}{\tau_\mu}\frac{\rme^{t^\prime/\tau_\mu}}{\rme^{{\rm min}(s,t)/\tau_\mu}-1},
\label{eq:modelA}
\\[1mm]
\mbox{Model B:}&~~& \beta^B_\mu(t,t^\prime\!,s)=\frac{a_\mu}{\tau_\mu}\rme^{-(t-t^\prime)/\tau_\mu}+\frac{a_\mu}{{\rm min}(s,t)}\Big\{1-\rme^{-t/\tau_\mu}\Big(\rme^{{\rm min}(s,t)/\tau_\mu}-1\Big)\Big\}.
\label{eq:modelB}
\end{eqnarray}
Both models are built around the time-translation invariant factor $\exp\left[-(t-t^\prime)/\tau_\mu\right]$ and satisfy  conditions (i), (ii), and (iii). So both reproduce the standard Cox model for time-independent covariates, as well as the instantaneous Cox model for longitudinal covariates with vanishing impact time scales, but they achieve this in distinct ways. We  could have ensured a time-translation invariant kernel $\beta_\mu(t,t^\prime\!,s)$ by choosing in Equation~(\ref{eq:general_form}) expressions for  $A_\mu(t,s)$ and $B_\mu(t,s)$ that are   independent of $t$. However, our  models would then not reduce to the standard Cox model when covariates are constant. 
For $t>s$ we find that $\beta^{\rm A}_\mu(t,t^\prime\!,s)$ is independent of $t$. This describes an  anomalous response: the system `remembers' early changes in covariates without decay. This could describe e.g. irreversible damage to the organism.  In contrast, $\beta^{\rm B}_\mu(t,t^\prime\!,s)$ retains a decaying dependence on $t$ when $t>s$, with  $\lim_{t \to\infty} \beta^{\rm B}_\mu(t,t^\prime\!,s)=a_\mu/s$.  This could describe, for example, fluctuations in hormone levels that impact the hazard mostly in the short term, but also with persistent long-term effects. 

Equations (\ref{eq:modelA}, \ref{eq:modelB}) only hold for $s>0$. In the data sets we study below there are some individuals whose longitudinal covariates are observed only once at the baseline time (i.e. their final observation time is $s=0$). Given that Equations (\ref{eq:modelA}) and (\ref{eq:modelB}) cannot be used for such individuals, we must specify their association parameters $\beta_\mu(t)$ in some other way. Two possible options are a constant association, $\beta_\mu(t)=a_\mu$, or a decaying association, $\beta_\mu(t)=a_\mu  \rme^{-t/\tau_\mu}$. Throughout the main paper we choose the former in the retarded kernel models. Results for the decaying association are presented in Section~S5 of the Supplementary Material. 

We note that we condition on knowledge of the covariates observed over a specific time interval $[0,s]$ in the model setup. As a consequence, the parameters in the retarded kernel models cannot necessarily be interpreted directly in terms of biophysical mechanisms. For example, $\tau_\mu$ encapsulates both the possible  decay in the physical effect of covariate $\mu$, and the decay that occurs from conditioning on knowledge of the covariate in the past. Parameter interpretations for the model therefore only make sense in a prediction context.

\subsubsection{Maximum likelihood inference.}
\label{Max_Like}
As in the standard Cox model, we use maximum likelihood inference to determine the most plausible values of the model parameters based on the observed data. For simplicity, in this section we will mostly omit the superscript RK from the hazard function. We write $\theta$ for the full set of parameters, i.e. the model parameters $\{\tau_\mu,a_\mu\}$ described in Section~\ref{sec:simple} and the base hazard rate $h_0(t)$, and assume initially that for each sample $i$ the covariates are known over the full time interval $[0,s_i]$.  The optimal parameters are those for which the data likelihood $\mathcal{P}(\mathcal{D}|\theta)$ is maximised. For non-censored data  this likelihood is given by  
\begin{equation}
    \mathcal{P}(\mathcal{D}|\theta) = \prod_{i=1}^{N} p(T_i|\theta, \mathcal{Z}^i_{[0,s_i]}), 
\end{equation}
where $p(t|\theta,  \mathcal{Z}^i_{[0,s_i]})$ is the probability density for individual $i$ experiencing an event at time $t$ given their covariate measurements. This probability density is expressed in terms of the parameterised hazard rate $h(t|\theta, \mathcal{Z}^i_{[0,s_i]})$ and the survival probability $S(t|\theta, \mathcal{Z}^i_{[0,s_i]})= \exp[-\int_0^t \rmd t^\prime h(t^\prime|\theta, \mathcal{Z}^i_{[0,s_i]})]$ via
\begin{align}
\label{eq:p_dist}
    p(t|\theta,  \mathcal{Z}^i_{[0,s_i]}) = h(t|\theta, \mathcal{Z}^i_{[0,s_i]})S(t|\theta, \mathcal{Z}^i_{[0,s_i]}).
\end{align}
For right-censored data there are two contributions to the likelihood. Individuals for whom an event is observed at time $T_i=T_i^*$ contribute a density $p(T_i|\theta, \mathcal{Z}^i_{[0,s_i]})$. Those that are censored at time $T_i=C_i$ contribute the survival probability $S(T_i|\theta, \mathcal{Z}^i_{[0,s_i]})$. Using the primary event indicator $\delta_i=I(T_i^*\leq C_i)\in\{0,1\}$, the likelihood for censored data is then
\begin{align}
    \mathcal{P}(\mathcal{D}|\theta) &= \prod_{i=1}^{N}  h(T_i|\theta, \mathcal{Z}^i_{[0,s_i]})^{\delta_i}S(T_i|\theta, \mathcal{Z}^i_{[0,s_i]}) .
\end{align}
Upon  defining $\Omega_{\mathrm{ML}} (\theta) = - \log \mathcal{P}(\mathcal{D}|\theta)$, we can write the maximum likelihood parameter estimators as $\hat{\theta}_{\text{ML}} = \text{argmin}_{\theta}\Omega_{\text{ML}}(\theta)$. 

A full derivation of the maximum likelihood equations for models of the form in Equation~(\ref{eq:retkernel}) is provided in Section~S1.1 of the Supplementary Material. Here we present only the results. The maximum likelihood estimator of the base hazard rate is the  direct analogue of the Breslow estimator:\cite{Breslow:1972}
\begin{align}
\label{eq:BaseHaz}
    \hat{h}_0 (t) =  \frac{\sum_{i=1}^{N} {\delta_i} \delta(t-T_i)}{\sum_{i=1}^{N} I(t\in[0,T_i])  \rme^{\sum_\mu \int_0^{\min(s_i,t)} \beta_\mu (t,t^\prime,s_i)z^i_\mu(t^\prime) \rmd t^\prime } },
\end{align}
recalling from Section~\ref{SetUp} that $I(A)=1$ if $A$ holds, and  $I(A)=0$ otherwise. The remaining parameters $\{a_\mu,\tau_\mu\}$ in Equations (\ref{eq:modelA}) and (\ref{eq:modelB}) are found by minimisation of 
\begin{eqnarray}
\label{eq:Omega}
    \Omega_{\mathrm{ML}}[\{a_\mu,\tau_\mu\}] &=& \sum_{i=1}^{N} \delta_i \log \Bigg( \sum_{j=1}^{N} I(T_i \in [0,T_j])  \rme^{\sum_\mu\int_0^{\min(s_j,T_i)} \beta_\mu (T_i,t^\prime,s_j)z^j_\mu(t^\prime) \rmd t^\prime  } \Bigg) \nonumber
     \\
    &&\hspace*{30mm} - \sum_{i=1}^{N} \delta_i \sum_\mu\int_0^{s_i}  \!\!\beta_\mu (T_i,t^\prime,s_i)z^i_\mu(t^\prime)\rmd t^\prime ,
\end{eqnarray}
where we have disregarded terms that are independent of $\{a_\mu,\tau_\mu\}$. As in all Cox-type models,  the final  minimisation of Equation~(\ref{eq:Omega}) with respect to the remaining parameters (here, the associations and time-scales) must be performed numerically, for example using Powell's method.\cite{Powell:1964}

\subsubsection{Dynamic Prediction.}
\label{OurModel_Pred}
Using the maximum likelihood estimates $\hat\theta_{\mathrm{ML}}$ for the model parameters, we can use the retarded kernel models to estimate the quantity $\pi(u|\mathcal{Z}_{[0,s]},s)$ in  Equation~(\ref{eq:Pi-1}), representing the probability that a subject has not experienced an event by time $u>s$,  conditional on their  survival to $s$ and on their covariate values $\mathcal{Z}_{[0,s]}$ up to that time. That is, 
\begin{align}
\label{eq:Pi_int}
    \hat{\pi}^{\mathrm{RK}}(u|\mathcal{Z}_{[0,s]},s) = \exp\Bigg\{-\int_{s}^{u} \hat{h}^{\mathrm{RK}}(t^\prime|\mathcal{Z}_{[0,s]})  \rmd t^\prime\Bigg\}, 
\end{align}
with $\hat{h}^{\mathrm{RK}}(t|\mathcal{Z}_{[0,s]})$ as defined by Equation~(\ref{eq:retkernel}), with kernels of the form in Equations (\ref{eq:modelA}, \ref{eq:modelB}) and with the ML estimators for the parameters in those kernels.  Using the ML estimator in Equation~(\ref{eq:BaseHaz}) of the base hazard rate  we can perform the integration in Equation~(\ref{eq:Pi_int}) to find 
\begin{align}
    \hat{\pi}^{\mathrm{RK}}(u|\mathcal{Z}_{[0,s]},s)\!= \!\exp\Bigg\{\!-\!\sum_{j=1}^{N} 
  \delta_j ~ I(T_j\!\in\![s,u])
        \frac{\rme^{\sum_{\mu=1}^{p} \int_0^{s}\hat\beta_\mu(T_j,t^\prime\!,s)z_\mu(t^\prime)\rmd t^\prime} }{\sum_{k=1}^{N} I(T_j \in [0,T_k])  \rme^{\sum_\mu \int_0^{\min(s_k,T_j)} \hat\beta_\mu (T_j,t^\prime,s_k)z^k_\mu(t^\prime) \rmd t^\prime } }   
    \Bigg\},
    \label{eq:predictor}
\end{align}
where $\hat{\beta}_\mu(t,t',s)$ indicates the association kernel obtained from the ML estimators of the parameters $\{a_\mu, \tau_\mu\}$. In the numerator we have used the fact that the prefactor $I(T_j\!\in \![s, u])$ ensures that $\min(s,T_j)\!=\!s$. 

\subsubsection{Covariate interpolation.}
So far, we have defined the retarded kernel models conditional on covariate trajectories $\mathcal{Z}_{[0,s]}$ over the entire interval $[0,s]$. In reality, we do not have full knowledge of these trajectories. Instead for each subject $i$ we have a finite number of discrete measurements that coincide with follow up appointments, $\mathcal{Z}^i=\{z^i_\mu(t_{i\ell}); ~\mu=1,\hdots,p, \ell=1,\hdots,n_i,  t_{i\ell} \in [0,s_i]\}$. In order to perform the integrals in  Equations (\ref{eq:Omega}) and (\ref{eq:predictor}) we must interpolate between these discrete observed values.

\begin{figure}[h]
    \centering
    \includegraphics[width=0.65\linewidth]{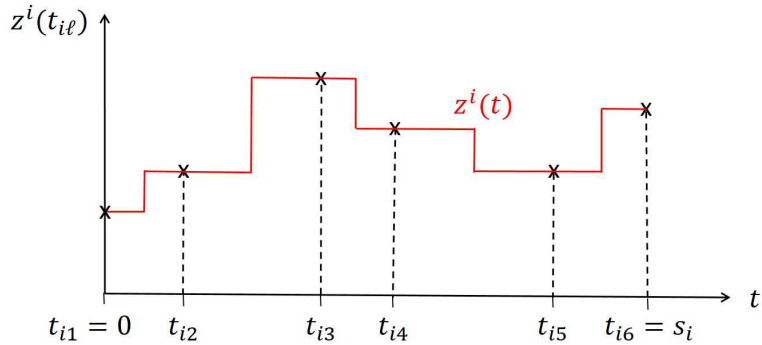}
    \caption{ An illustration of the interpolation method for covariates. For each subject $i$, there are a discrete number of covariate observations. The observation times $t_{i\ell}$ are labelled on the horizontal axis. The covariate measurement at each observation time is indicated by a cross. The solid line shows the interpolated covariate trajectory based on these discrete observations. The value of a covariate at time $t \neq t_{i\ell}$ is taken to be equal to the observed value of the covariate at the observation time closest to $t$. This yields a step function that changes value half way between each pair of consecutive observations. }
    \label{fig:StepFunc}
\end{figure}

We choose a simple `nearest neighbour' approach, that is we set $z^i_\mu(t)=z^i_\mu(t_{i\ell})$ where $t_{i\ell}$ is the observation time closest to $t$. The approximate covariate trajectory is then a step function that changes value half way between each pair of consecutive observation times. Figure~\ref{fig:StepFunc} illustrates this idea. Using this method, the integrals in  Equations (\ref{eq:Omega}) and (\ref{eq:predictor}) can be evaluated analytically (see Section~S1.3 of the Supplementary Material). This reduces the computational effort required to perform the minimisation and the dynamic prediction. Other, smoother interpolation procedures such as Gaussian convolutions\cite{Press:1992, VanDenBoomgaard:2001} are also possible and may improve estimations (at some computational cost). While interpolation makes assumptions about the values of the covariate within the observation interval $[0,s_i]$, we do not make assumptions about the covariates after the final observation time $s_i$. 

\section{Application to clinical data}
\label{Apply}
\subsection{Methods}
\subsubsection{Training and test data.} For each of the data sets we assess the predictive accuracy of the different dynamic prediction models by splitting the data in half into a training data set and a test data set. Each model is fitted to the training data and the resulting model is used to make survival predictions about individuals in the test data set. Predictive accuracy is assessed by comparing these predictions to the true event times in the test data (see Section~\ref{PredError}). The procedure was repeated for 20 random splits of the data and the corresponding prediction error was averaged over these repetitions. 

\subsubsection{Measuring predictive accuracy.}
\label{PredError}
Following Rizopoulos et al. (2017),\cite{Rizopoulos:2017} we quantify the predictive accuracy of the different models using the expected error of predicting future events. Dynamic prediction is concerned with predicting the survival of individuals to a given time $u$, based on their survival to some earlier time $t<u$, and covariate measurements for the individual up to this time. The expected prediction error for a given `prediction time' $u$ and `base time' $t$ is then defined as\cite{Schoop:2011}
\begin{align}
    \mathrm{PE}(u|t) = \mathbb{E}\left[ L\{ N_i(u) - \pi(u|\mathcal{Z}^i_{[0,t]}, t) \} \right]
\end{align} 
where $N_i(u)=I(T_i^*\!>u)$ is the true event status of subject $i$ at time $u$, and    $\pi(u|\mathcal{Z}^i_{[0,t]}, t)$ is the model's predicted survival probability for subject $i$ based on information about this subject (covariate measurements and survival status) up to the base time $t$. The notation $\mathbb{E}$ stands for an average over the distribution of covariates and event times. $L(.)$ denotes a loss function which defines how we measure the difference between survival status and predicted survival probability. Commonly choices are  $L(x)=|x|$ and the squared loss $L(x)=x^2$.\cite{Rizopoulos:2017, Henderson:2002, Schoop:2011} We choose the latter. The definition of prediction error is such that $\mathrm{PE}(u|t)=0$ if the survival status of all individuals is predicted with full accuracy (i.e. $\pi(u|\mathcal{Z}^i_{[0,t]}, t)=1$ for all subjects who are alive at time $u$ and $\pi(u|\mathcal{Z}^i_{[0,t]}, t)=0$ for subjects who are dead by time $u$). If the reverse is true ($\pi(u|\mathcal{Z}^i_{[0,t]}, t)=1$ for subjects who are dead at time $u$ and $\pi(u|\mathcal{Z}^i_{[0,t]}, t)=0$ for subjects who are alive) then $\mathrm{PE}(u|t)=1$. We obtain $\mathrm{PE}(u|t)=0.25$ if every individual has predicted survival probability $\pi(u|\mathcal{Z}^i_{[0,t]}, t)=0.5$.

Again following Rizopoulos et al. (2017),\cite{Rizopoulos:2017} in this paper we use the overall prediction error $\mathrm{PE}(u|t)$ proposed by Henderson et al. (2002),\cite{Henderson:2002} that in addition takes into account censoring,
\begin{multline}
\label{eq:PE-hat}
    \widehat{\mathrm{PE}}(u|t) = \frac{1}{n(t)} \sum_{i; T_i\geq t} I(T_i \geq u) L\{1-\hat{\pi}(u|\mathcal{Z}^i_{[0,t]}, t)\} + \delta_i I(T_i<u) L\{0-\hat{\pi}(u|\mathcal{Z}^i_{[0,t]}, t)\} + \\
    (1-\delta_i)I(T_i<u)\Big[ \hat{\pi}(u|\mathcal{Z}^i_{[0,t]}, T_i)L\{ 1 - \hat{\pi}(u|\mathcal{Z}^i_{[0,t]}, t)\} + \{1-\hat{\pi}(u|\mathcal{Z}^i_{[0,t]}, T_i)\}L\{0-\hat{\pi}(u|\mathcal{Z}^i_{[0,t]}, t)\} \Big].
\end{multline} 
The sum extends over the $n(t)$ subjects in the test data set who are still at risk at time $t$. The first term of Equation~(\ref{eq:PE-hat}) corresponds to individuals in the test data who are still alive after time $u$. These have survival status $N_i(u)=1$, and therefore contribute a loss function based on the difference between their estimated survival probability and 1, i.e., $L\{1-\hat{\pi}(u|\mathcal{Z}^i_{[0,t]}, t)\}$. The second term refers to individuals who have experienced an event by time $u$ (i.e. $T_i=T_i^*<u$). Their survival status is 0 and therefore they contribute a loss function $L\{0-\hat{\pi}(u|\mathcal{Z}^i_{[0,t]}, t)\}$. The final term represents individuals who were censored before time $u$ (i.e. $T_i=C_i \in [t,u]$) so we do not know their survival status at time $u$. Here the estimated probability of survival based on information up to time $t$ is compared with the probability of survival given that we know subject $i$ survived up until their censoring time $T_i\geq t$. 

To compare the predictive accuracy of joint modelling, landmarking and the retarded kernel approach we insert into Equation~(\ref{eq:PE-hat}) the respective estimators $\hat{\pi}^{\mathrm{JM}}(u|\mathcal{Z}^i_{[0,t]}, t)$, $\hat{\pi}^{\mathrm{LM}}(u|\mathcal{Z}^i_{[0,t]}, t)$ and $\hat{\pi}^{\mathrm{RK}}(u|\mathcal{Z}^i_{[0,t]}, t)$.  This requires that we  calculate the probability of a subject's survival to time $u$, based on survival and covariate observations until a general base time $t<u$ that need not be the individual's final observation time $s_i$. For the joint model and landmarking estimators we replace the final observation time with $t$ in Equations (\ref{eq:Pi-JM}) and (\ref{eq:LM_pi}). For the retarded kernel estimator $\hat{\pi}^{\mathrm{RK}}(u|\mathcal{Z}^i_{[0,t]}, t)$ in Equation~(\ref{eq:predictor}) we replace $I(T_j\in[s_i,u])$ with $I(T_j\in[t,u])$ since we know subject $i$ is alive until $t$. However, we only have covariate observations up to the latest observation time $t_{i\ell}$ that is $\leq t$. In line with our chosen interpolation procedure we only integrate the covariate trajectory up to this time. Specifically, for any general base time $t$ we have 
\begin{align}
\label{eq:Pi-Ours-gen}
    \hat{\pi}^{\mathrm{RK}}(u|\mathcal{Z}^i_{[0,t]}, t)\!=\!\exp\left\{\!-\!\sum_{j=1}^{N} \delta_j~ I( T_j\! \in\! [t, u]) \frac{ \rme^{ \sum_\mu\int_0^{\max\{t_{i\ell}:t_{i\ell}\leq t\}}  \hat\beta_\mu(T_j,t^\prime\!,s_i) z^i_\mu(t^\prime)\rmd t^\prime }}{\sum_{k=1}^{N} I(T_j \in [0,T_k])  \rme^{ \sum_\mu 
    \int_0^{\min(s_k,T_j)}\hat\beta_\mu(T_j,t^\prime,s_k) z^k_\mu(t^\prime) \rmd t^\prime } }  \right\},
\end{align}
where index $i$ labels the individual (in the test data) for whom we are making predictions, while the sums over $j$ and $k$ refer to individuals in the training data set used for inference. The integral limit $\max\{t_{i\ell}:t_{i\ell}\leq t\}$ labels the last observation time of individual $i$ before (or at) the base time $t$.  

The term $\hat{\pi}(u|\mathcal{Z}^i_{[0,t]}, T_i)$ in Equation~(\ref{eq:PE-hat}) represents the probability of survival to $u$ given subject $i$ survived to their censoring time $T_i=C_i$. To calculate this using the retarded kernel model we replace $I( T_j\! \in\! [t, u])$ with $ I( T_j\! \in\! [T_i, u])$ in Equation~(\ref{eq:Pi-Ours-gen}). For joint modelling $\hat{\pi}^{\mathrm{JM}}(u|\mathcal{Z}^i_{[0,t]}, T_i)$ is obtained by replacing $\mathcal{Z}_{[0,s]}$ with $\mathcal{Z}^i_{[0,t]}$ and by replacing the condition $T^*>s$ with $T_i^*>T_i$ in  Equation~(\ref{eq:Pi-JM}). Since this term is only calculated for censored individuals ($T_i=C_i$), the condition $T_i^*>T_i$ means `the true event time of individual $i$ is greater than their censoring time'. Finally, for landmarking we use $\hat{\pi}^{\mathrm{LM}}(u|\mathcal{Z}^i_{[0,t]}, T_i) = \hat{\pi}^{\mathrm{LM}}(u|\mathcal{Z}^i_{[0,t]}, t)/\hat{\pi}^{\mathrm{LM}}(T_i|\mathcal{Z}^i_{[0,t]}, t)$ which is equivalent to replacing $s$ with $t$ in Equation~(\ref{eq:LM_pi}) except in the integral limits where we replace $\int_s^u$ with $\int_{T_i}^u$. 

To perform the prediction error calculation for the retarded kernel models we use our own \texttt{C++} code following Equations (\ref{eq:PE-hat}) and (\ref{eq:Pi-Ours-gen}). For the joint model and landmarking model we use a version of the function \texttt{prederrJM} in the \texttt{JMbayes} package subject to minor modifications (see Section~S4 of the Supplementary material for details).

\subsubsection{Fixed base time.}
First we compare the predictive accuracy of the three methods by specifying a fixed base time $t$ and varying the prediction time $u$. Based on Figures~\ref{fig:Mayo} and \ref{fig:Proth}, for the PBC and Liver data sets we choose a fixed base time of $t=3$ years. This value is chosen so that a large number of individuals are still alive after this time (and we can hence make predictions about them), but also so that these individuals have had their covariates measured multiple times before this time. We then vary the prediction time $u$ from the base time $t=3$ years in steps of $0.2$ years up to $8$ years for the PBC data, and up to $10$ years for the Liver data. For the AIDS data set we choose $t=6$ months as the base time, so that most individuals have been observed three times. We then vary the prediction time $u$ from this base time up to $18$ months in steps of $0.2$ months. 

\subsubsection{Fixed prediction window.}
In our second test, we vary the base time $t$, while keeping the prediction window $w=u-t$ fixed (i.e., the time difference between prediction and  base time). Since we are varying the base time $t$, we must then fit a new landmark model for each choice of $t$ (where the landmark time $\upsilon=t$). On the other hand, for the retarded kernel approach and the joint model we need only fit the model once, and can make the error assessments at each iteration using this single fitted model.

Based on previous analysis of the PBC and Liver data\cite{Rizo:2012, Henderson:2002} we choose three prediction windows: $w_1=1$ year, $w_2=2$ years, and $w_3=3$ years. Given the event time distributions, we do not make predictions for either data set beyond $u=10$ years. Therefore, for $w_1$ we vary the base time from $0-9$ years, for $w_2$ we vary it from $0-8$ years and for $w_3$ this is $0-7$ years. In all cases we increase the base time in increments of $0.2$ years. 

Based on the event time distribution of the AIDS data, we choose prediction windows $w_1=6$ months, $w_2=9$ months and $w_3=12$ months. Here covariates are observed at 0, 2, 6, 12, and 18 months only. As a result, predictions will only be updated at these time steps, and we can only make a small number of distinct measurements of predictive accuracy. Due to the event times in the AIDS data set, we do not make predictions past 18 months. Therefore, for window $w_1$ we use base times $t=0,2,6,12$ months and for windows $w_2$ and $w_3$ we use $t=0,2,6$ months only.

\subsection{PBC data set}
\label{PBC_results}
We fit each model to the PBC training data set using $p=3$ time-dependent covariates and a single fixed covariate: $z^i_1(t)$ denotes log serum bilirubin, $z^i_2(t)$ denotes log serum albumin, $z^i_3(t)$ is log prothrombin time,  and the fixed covariate $\zeta^i_1$ is the subject's age at baseline. 

The PBC data set contains event-time information for two events, death and liver transplant. The most appropriate way of analysing this data is to use a competing risks model. However, for simplicity we here treat the transplant event as a censoring event. Another simple way to analyse this data is to treat the two events as a single composite event. We provide the results of the latter analysis in Section~S3 of the Supplementary Material. The two analyses are found to give similar results. 

\subsubsection{Models.}
For the joint model we first fit a simple multivariate linear mixed model to each of the three time-dependent covariates,
\begin{align}
    z^i_\mu (t) = m^i_\mu(t) + \varepsilon^i_\mu (t) = \eta_{\mu,0} + b^i_{\mu,0} + (\eta_{\mu,1} + b^i_{\mu,1})t + \varepsilon^i_\mu (t),
\end{align}
where the random effects $\boldsymbol{b}^i$ are assumed to follow a joint multivariate normal distribution  with mean zero and  variance-covariance matrix $\boldsymbol{D}$.

Figure~\ref{fig:Mayo} suggests that the covariate trajectories in the PBC data may be non-linear for some individuals. Hence, for extra flexibility we also fit a second joint model that includes natural cubic splines in both the fixed and random effects parts of the model. Following Rizopoulos (2016),\cite{Rizo:2016} the log serum bilirubin ($\mu=1$) is modelled using natural cubic B-splines with 2 degrees of freedom,
\begin{align}
    z^i_1(t) &= m^i_1(t) + \varepsilon^i_1 (t) \nonumber \\
    &=  \eta_{1,0} + b^i_{1,0} + (\eta_{1,1} + b^i_{1,1}) B_n (t,\lambda_{1}) + (\eta_{1,2} + b^i_{1,2})B_n(t,\lambda_2) + \varepsilon^i_1 (t) 
\end{align}
where $\{B_n(t,\lambda_k);k=1,2\}$ denotes the B-spline basis matrix for a natural cubic spline of time.\cite{Rizo:2012, Perperoglou:2019} We write analogous equations for both the log albumin and the log prothrombin covariates. Again, the random effects of all three longitudinal covariates are assumed to follow a joint multivariate normal distribution. 

For both the linear and spline longitudinal models, the hazard function of the survival sub-model in the joint modelling framework is
\begin{align}
     h^{\mathrm{JM}} (t|\mathcal{M}^i_{[0,t]}) = h_0(t) \exp \left\{ \gamma_1 \zeta^i_1 + \alpha_1 m^i_1(t) + \alpha_2 m^i_2 (t) + \alpha_3 m^i_3 (t) \right\},
\end{align}
where we recall from Section~\ref{JointModels} that $\mathcal{M}^i_{[0,t]}=\{m^i_\mu(t'); ~\mu=1,\hdots,p, ~t'\in[0,t]\}$ denotes the history of the `true' (unobserved) longitudinal covariates up to time $t$ for subject $i$. For the landmark model the hazard is instead specified for a given landmark time, $\upsilon$,
\begin{align}
    h^{\mathrm{LM}}(t | \mathcal{Z}^i,\upsilon) = h_0 (t|\upsilon) \exp\left\{ \gamma_1 \zeta^i_1 + \alpha_1(\upsilon) \tilde{z}^i_1(\upsilon) + \alpha_2(\upsilon) \tilde{z}^i_2(\upsilon) + \alpha_3(\upsilon) \tilde{z}^i_3(\upsilon) \right\},
\end{align}
where $\tilde z_\mu^i(\upsilon)$ is again the last observed value of covariate $\mu$ for patient $i$ before time $\upsilon$. 

For the retarded kernel approach, we specify the hazard function as
\begin{multline}
    h^{\mathrm{RK}} (t|\mathcal{Z}^i_{[0,s_i]}) = h_0(t) \exp \Bigg\{ \gamma_1 \zeta^i_1 + \int_0^{\min(s_i,t)} \left( \beta_1(t,t^\prime,s_i) z^i_1(t^\prime) + \beta_2(t,t^\prime,s_i) z^i_2(t^\prime) \right.  \\[-1mm] \left.  + \beta_3(t,t^\prime,s_i) z^i_3(t^\prime)\right) \rmd t^\prime\Bigg\}.
\end{multline}
The parameterisations of the time-dependent association parameters $\beta_\mu(t,t',s)$ are given in Equations (\ref{eq:modelA}) and (\ref{eq:modelB}) for models A and B, respectively.

\subsubsection{Results.}
\begin{figure}[h]
    \centering
    \includegraphics[width=0.6\linewidth]{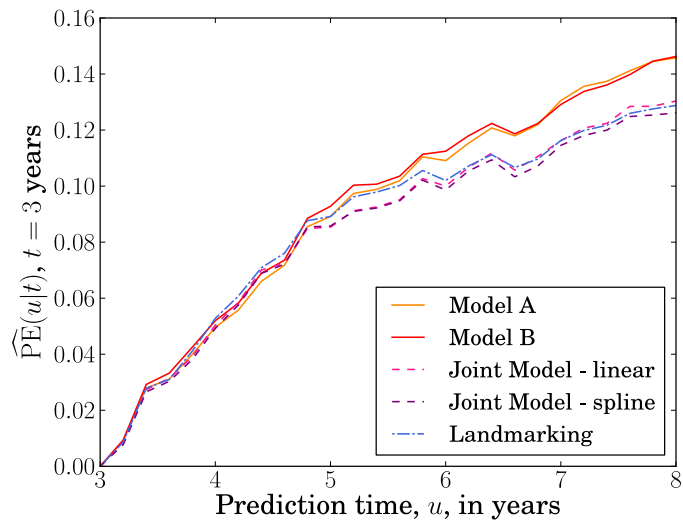}
    \caption{~Overall prediction error $\widehat{\mathrm{PE}}(u|t)$ as a function of prediction time $u$ (in years) for the PBC data with fixed base time $t=3$ years.  Prediction error is calculated for $u$ values from 3 to 8 years, with 0.2 year increments. A squared loss function was used in Equation~(\ref{eq:PE-hat}). The prediction error plotted at each time $u$ is an average over values of $\widehat{\mathrm{PE}}(u|t)$ calculated for 20 random splits of the data into training and test data sets. The results from models A and B of the retarded kernel approach  are plotted alongside the landmarking model and two joint models (one that uses a linear longitudinal model for the time-dependent covariates, and another that uses cubic splines).}
    \label{fig:PE-PBC}
\end{figure}

Figure~\ref{fig:PE-PBC} shows plots of the overall prediction error $\widehat{\mathrm{PE}}(u|t)$ against the prediction time $u$ for a fixed base time of $t=3$ years averaged over the 20 random splits of the data. Results for the linear joint model, spline joint model, landmarking model and models A and B of the retarded kernel approach are plotted on the same graph. All five models have similarly accurate predictions up to $u=5$ years. For later prediction times, the standard approaches have a lower average prediction error than the retarded kernel models. The largest disparity in prediction error is observed  at $u=8$ years between the spline joint model ($\widehat{\mathrm{PE}}(u|t)=0.126$) and the retarded kernel models which both have $\widehat{\mathrm{PE}}(u|t)=0.146$. 

\begin{figure}[h]
    \centering
    \includegraphics[width=1.0\linewidth]{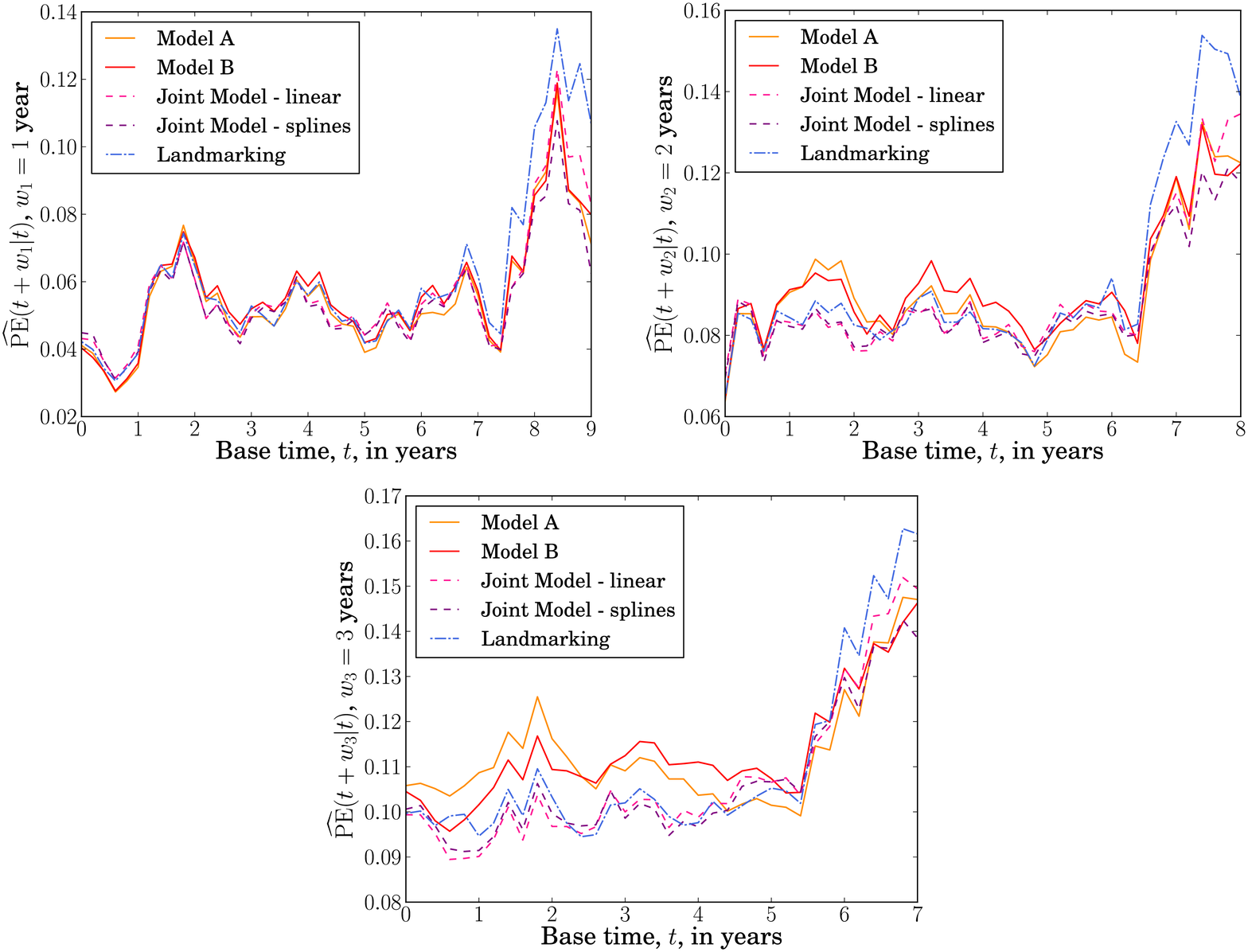}
    \caption{~ Overall prediction error $\widehat{\mathrm{PE}}(u|t)$ versus base time $t$ (in years) for the PBC data, with prediction windows $w_1=1$ year, $w_2=2$ years and $w_3=3$ years. The prediction times are $u=t+w$. The prediction error is calculated for $t$ ranging from 0 to 9,8 or 7 years for $w_1$, $w_2$ and $w_3$ respectively, with 0.2 year increments. A squared loss function was used in Equation~(\ref{eq:PE-hat}). The prediction error plotted at each time $t$ is an average over values of $\widehat{\mathrm{PE}}(u|t)$ calculated for 20 random splits of the data into training and test data sets. Results from models A and B of the retarded kernel approach are plotted alongside the landmarking model  and two joint models; one that uses a linear longitudinal model for the time-dependent covariates, and another that uses cubic splines. }
    \label{fig:PW-PBC}
\end{figure}

Plots of the average prediction error $\widehat{\mathrm{PE}}(u|t)$ against the base time $t$ are shown in Figure~\ref{fig:PW-PBC}, for fixed prediction windows $w_1=1$ year, $w_2=2$ years, and $w_3=3$ years. Again results for the five different models are plotted in the same graphs. For the shortest prediction window $w_1$, all  models are similarly accurate for base times up to $t=7.5$ years, after which the landmarking model performs slightly worse than the others. For the larger prediction windows, models A and B of the retarded kernel approach show slightly larger prediction errors than the other models over the range $t=0-5$ years. At larger base times the joint models and retarded kernel models again exhibit similar prediction errors while landmarking has the largest error. The largest difference in performance occurs at $t=9$ years for the prediction window $w_1$, between the spline joint model ($\widehat{\mathrm{PE}}(u|t)=0.062$) and the landmarking model ($\widehat{\mathrm{PE}}(u|t)=0.107$). 

The results of the above tests suggest that, for the PBC data, the retarded kernel approach performs as well as existing methods for prediction windows $<2$ years but less well for larger windows. However, as the base time increases, the retarded kernel models behave similarly to the joint models while the landmarking model exhibits the highest prediction error. Care should be taken when interpreting these results, as we have not used a competing risks model in our analysis. This data set does, however, serve as an illustration that with only a modest drop in accuracy the retarded kernel model can serve as a simpler alternative to joint modelling when considering multiple longitudinal covariates. Unlike the landmarking approach, the retarded kernel model takes into account the full history of covariate observations which, along with the fact that landmarking discards more data as the base time increases, may explain why the landmarking model performs worst for later base times.  

\subsection{AIDS data set}
In the AIDS data set we focus on a single longitudinal covariate, the CD4 count $z^i_1(t)$. We also include four fixed binary covariates: drug group  ($\zeta^i_1=1$ for ddI and $\zeta^i_1=0$ for ddC), gender ($\zeta^i_2=1$ for male and $\zeta^i_2=0$ for female), PrevOI \ ($\zeta^i_3=1$ for AIDS diagnosis at study entry and $\zeta^i_3=0$ for no AIDS diagnosis) and Stratum  ($\zeta^i_4=1$ for AZT failure and $\zeta^i_4=0$ for AZT intolerance). See Section~\ref{AIDS} for a description of these variables.

\subsubsection{Models.}
The joint modelling framework allows us to model the dependence of CD4 count on the patients drug group. Following Rizopoulos (2012)\cite{Rizo:2012} we fit the linear mixed model,
\begin{align}
    z^i_1(t) &= m^i_1(t) + \varepsilon^i_1(t) \nonumber\\
    &=\eta_{1,0} + b^i_{1,0} + (\eta_{1,1} + b^i_{1,1})t + \eta_{1,2}  \zeta^i_1 t + \varepsilon^i_1 (t),
\end{align}
where the term $\eta_{1,2}  \zeta^i_1 t $ denotes the effect of the interaction of treatment (drug group) with time. As usual, the random effects $\boldsymbol{b}^i$ are assumed to follow a normal distribution. 
To complete the joint model, the hazard function is then chosen as
\begin{align}
    h^{\mathrm{JM}} (t|\mathcal{M}^i_{[0,t]}) = h_0(t) \exp \left\{ \gamma_1 \zeta^i_1 +  \gamma_2 \zeta^i_2 +  \gamma_3 \zeta^i_3 +  \gamma_4 \zeta^i_4 + \alpha_1 m^i_1 (t)  \right\}.
\end{align}
For the landmark model with landmark time $\upsilon$ one has 
\begin{align}
    h^{\mathrm{LM}}(t | \mathcal{Z}^i,\upsilon) = h_0 (t|\upsilon) \exp\left\{ \gamma_1 \zeta^i_1 + \gamma_2 \zeta^i_2 +  \gamma_3 \zeta^i_3 +  \gamma_4 \zeta^i_4 + \alpha_1(\upsilon) \tilde{z}^i_1(\upsilon)  \right\},
\end{align}
and for the retarded kernel approach we specify the survival model as follows
\begin{align}
    h^{\mathrm{RK}} (t|\mathcal{Z}^i_{[0,s_i]}) = h_0(t) \exp \Bigg\{ \gamma_1 \zeta^i_1 +  \gamma_2 \zeta^i_2 +  \gamma_3 \zeta^i_3 +  \gamma_4 \zeta^i_4 + \int_0^{\min(s_i,t)}  \!\beta_1(t,t^\prime\!,s_i) z^i_1(t^\prime) \rmd t^\prime  \Bigg\}.
\end{align}
As before, the parameterisations of $\beta_\mu(t,t^\prime\!,s)$ in models A and B are given in Equations (\ref{eq:modelA}) and (\ref{eq:modelB}) respectively.

\begin{figure}[b]
    \centering
    \includegraphics[width=0.6\linewidth]{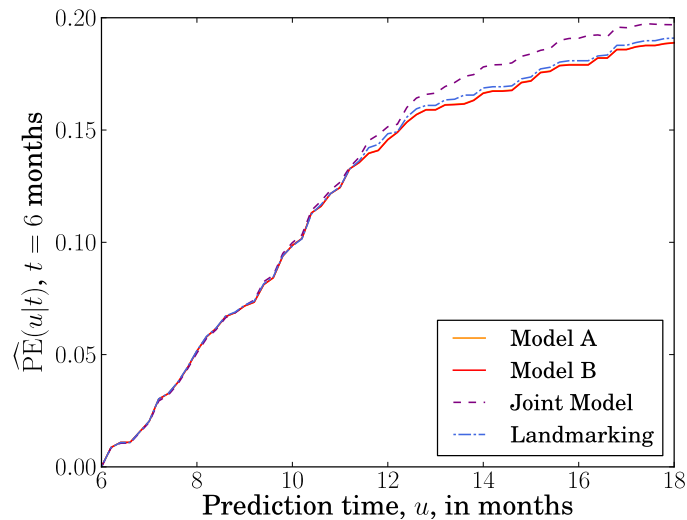}
    \caption{~ Overall prediction error $\widehat{\mathrm{PE}}(u|t)$ plotted versus prediction time $u$ (in months) for the AIDS data with fixed base time $t=6$ months. This error is calculated for $u$ ranging from 6 to 18 months, at  0.2 month intervals. In Equation~(\ref{eq:PE-hat}) a squared loss function was used. The prediction error plotted at each time $u$ is an average over values of $\widehat{\mathrm{PE}}(u|t)$ calculated for 20 random splits of the data into training and test data sets. The results from retarded kernel models A and B are plotted alongside the results from the landmarking model and a joint model. The results from model A cannot be seen because they overlap with the results from model B. }
    \label{fig:PE-AIDS}
\end{figure}

\begin{figure}[h]
    \centering
    \includegraphics[width=1.0\linewidth]{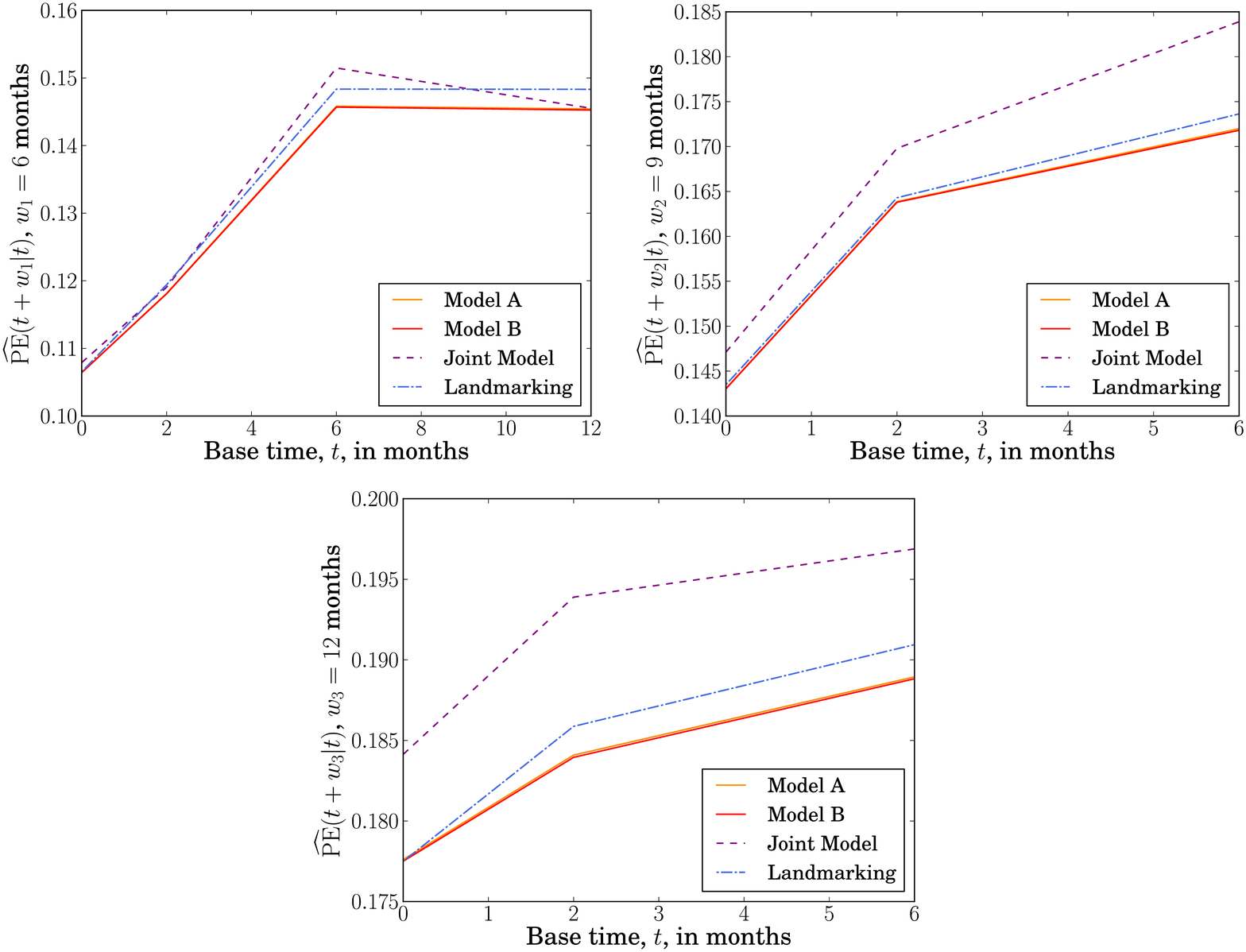}
    \caption{~ Overall prediction error $\widehat{\mathrm{PE}}(u|t)$ versus base time $t$ (in months) for the AIDS data with three fixed prediction windows: $w_1=6$ months, $w_2=9$ months and $w_3=12$ months. The prediction times are  $u=t+w$. Observations are made at times $0,2,6,12,18$ months for all individuals in this data set. Prediction errors are hence only updated at these time points. For prediction window $w_1$, prediction error is measured for $t=0,2,6$ and $12$ months. For windows $w_2$ and $w_3$, the error is measured at $t=0,2$ and $6$ months only. In Equation~(\ref{eq:PE-hat}) we used a squared loss function. The prediction error plotted at each time $t$ is an average over values of $\widehat{\mathrm{PE}}(u|t)$ calculated for 20 random splits of the data into training and test data sets. The results from retarded kernel models A and B are plotted alongside the landmarking model and a joint model. The results from model A cannot be seen clearly because they overlap with the results from model B.}
    \label{fig:PW-AIDS}
\end{figure}

\subsubsection{Results.}
The plots of $\widehat{\mathrm{PE}}(u|t)$ against prediction time $u$ with base time $t=6$ months are shown in Figure~\ref{fig:PE-AIDS} for the four models. As before, the data for $\widehat{\mathrm{PE}}(u|t)$ is an average over $20$ random splits of the data into training and test sets. All models show comparable accuracy up to $u=11$ months. After this time, the joint model shows slightly worse prediction than the other three (whose accuracies remain almost equal). The largest difference in predictive error occurs at $u=16.2$ months, between models A and B of the retarded kernel approach on the one hand and the joint model on the other. At this value of $u$, both versions of the retarded kernel model lead to $\widehat{\mathrm{PE}}(u|t)=0.179$ while the joint model has  $\widehat{\mathrm{PE}}(u|t)=0.192$.
%LM=0.181

Figure~\ref{fig:PW-AIDS} shows plots of  $\widehat{\mathrm{PE}}(u|t)$ against base time for the AIDS data set with three prediction windows, $w_1=6$ months, $w_2=9$ months and $w_3=12$ months. For the shortest prediction window $w_1$, all four models have similar prediction error at $t=0$ and $2$ months. The joint model has the largest error at $t=6$ months (where models A and B are lowest), but has the same error as the retarded kernel models at $t=12$ months (where landmarking has the highest error). For windows $w_2$ and $w_3$ the joint model demonstrates the worst prediction at all base times. The other three models exhibit similar errors at $t=0$ for both these windows as well as at $t=2$ for window $w_2$. In all other scenarios, models A and B of the retarded kernel approach have the lowest prediction error. The largest difference in prediction error is for $w_2$ at $t=6$ months where the joint model has $\widehat{\mathrm{PE}}(u|t)=0.184$, landmarking has $\widehat{\mathrm{PE}}(u|t)=0.174$ and the retarded kernel models both have $\widehat{\mathrm{PE}}(u|t)=0.172$.

The above results suggest that, for the AIDS data set, the joint model has the worst predictive accuracy overall, while the two retarded kernel models perform best.

\subsection{Liver data set}
For the liver data set we model prothrombin index as our one longitudinal covariate $z^i_1(t)$, and drug group as our single fixed covariate $\zeta^i_1$. The fixed covariate is defined such that $\zeta^i_1=1$ for individuals in the treatment (prednisone) group, and $\zeta^i_1=0$ for those in the placebo group.

\subsubsection{Models.} 

\begin{figure}[t]
    \centering
    \includegraphics[width=0.6\linewidth]{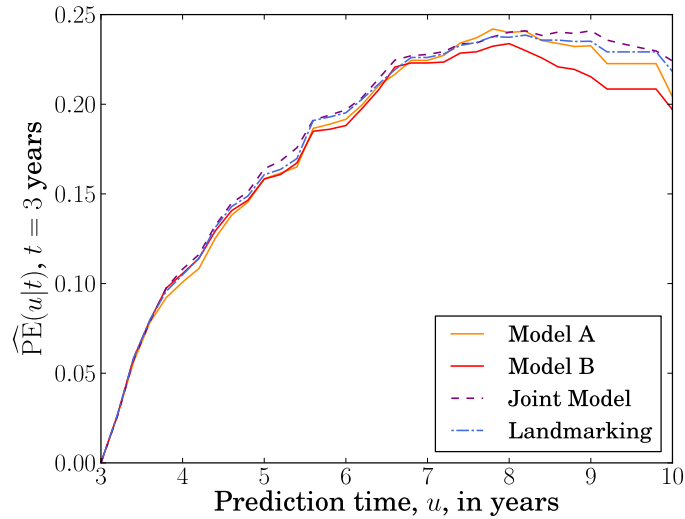}
    \caption{~ Overall  prediction error $\widehat{\mathrm{PE}}(u|t)$ plotted versus prediction time $u$ (in years) for the Liver data with fixed base time $t=3$ years. This error is calculated for $u$ ranging from 3 to 10 years, with 0.2 year increments.  In  Equation~(\ref{eq:PE-hat}) we used a squared loss function. The prediction error plotted at each time $u$ is an average over values of $\widehat{\mathrm{PE}}(u|t)$ calculated for 20 random splits of the data into training and test data sets. The results from retarded kernel models A and B are plotted alongside the results from the landmarking model and a joint model.}
    \label{fig:PE-Liver}
\end{figure}

Following Rizopoulos (2012),\cite{Rizo:2012} we define a flexible longitudinal model for the subject-specific prothrombin trajectories, using natural cubic splines with different average profiles for each drug group. Rizopoulos (2012)\cite{Rizo:2012} also suggests to include a separate indicator variable of the baseline measurement, to capture sudden changes in the prothrombin index in the early part of follow up. The longitudinal model then takes the form
\begin{eqnarray}  
z^i_1(t) &=&m^i_1(t) + \varepsilon_1^i(t) \nonumber\\
    &=& \eta_{1,0} + b^i_{1,0} + (\eta_{1,1} + b^i_{1,1}) B_n(t,\lambda_1) + (\eta_{1,2} + b^i_{1,2}) B_n(t,\lambda_2)  + (\eta_{1,3} + b^i_{1,3}) B_n(t,\lambda_3) \nonumber \\
    && + \eta_{1,4} \zeta^i_1 B_n(t,\lambda_1) + \eta_{1,5} \zeta^i_1 B_n(t,\lambda_2) + \eta_{1,6} \zeta^i_1 B_n(t,\lambda_3) + \eta_{1,7} \zeta^i_1 \nonumber\\
    && + \eta_{1,8} I(t=t_{i,1})  + \eta_{1,9} \zeta^i_1 I(t=t_{i,1})  + \varepsilon_i(t) 
\end{eqnarray}
where $I(t=t_{i,1})$ is the indicator variable for the baseline time and, as before, $\{B_n(t,\lambda_k); k=1,2,3\}$ is the B-spline matrix for a natural cubic spline of time. This time, two internal knots are placed at 33\% and 66.7\% percentiles of the follow up times. The random effects are assumed to have a diagonal covariance matrix. 

\begin{figure}[t]
    \centering
    \includegraphics[width=1.0\linewidth]{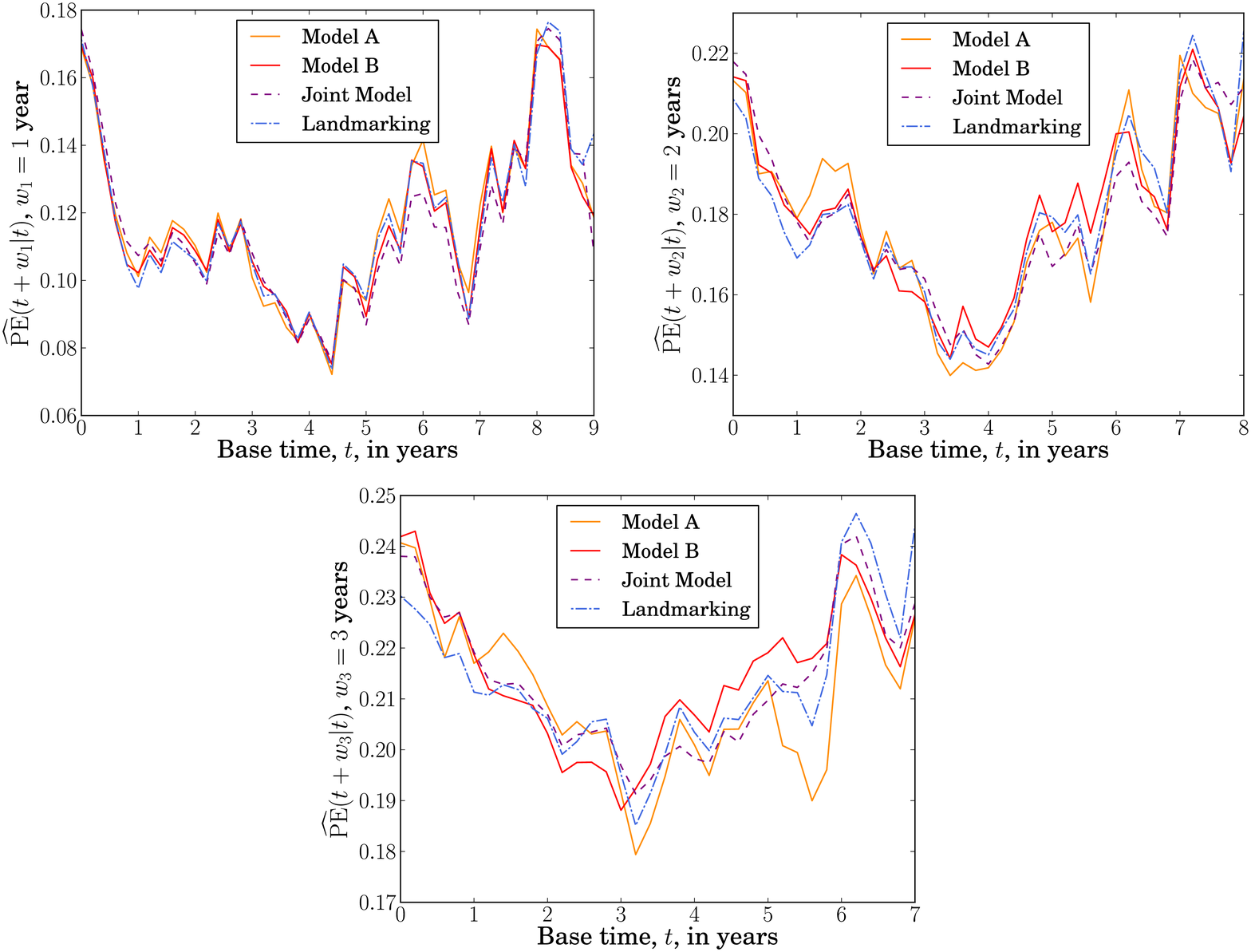}
    \caption{~ Overall prediction error $\widehat{\mathrm{PE}}(u|t)$ plotted against base time $t$ (in years) for the Liver data with three fixed prediction windows, $w_1=1$ year, $w_2=2$ years and $w_3=3$ years. The prediction times are $u=t+w$. The error is calculated for $t$ ranging from 0 to 9,8 or 7 years, for $w_1$, $w_2$ and $w_3$ respectively, with 0.2 year intervals.  In Equation~(\ref{eq:PE-hat}) a squared loss function was used. The prediction error plotted at each time $t$ is an average over values of $\widehat{\mathrm{PE}}(u|t)$ calculated for 20 random splits of the data into training and test data sets. The results from retarded kernel models A and B are plotted alongside the landmarking model and a joint model. }
    \label{fig:PW-Liver}
\end{figure}

The hazard functions for the joint model and the landmark model (with landmark time $\upsilon$) are then 
\begin{align}
    h^{\mathrm{JM}} (t|\mathcal{M}^i_{[0,t]}) &= h_0(t) \exp \left\{ \gamma_1 \zeta^i_1 +   \alpha_1 m^i_1(t)  \right\},\\
    h^{\mathrm{LM}}(t | \mathcal{Z}^i,\upsilon) &= h_0 (t|\upsilon) \exp\left\{ \gamma_1 \zeta^i_1 + \alpha_1(\upsilon) \tilde{z}^i_1(\upsilon)  \right\}.
\end{align}
For the retarded kernel models we have
\begin{align}
    h^{\mathrm{RK}} (t|\mathcal{Z}^i_{[0,s_i]}) = h_0(t) \exp \Bigg\{ \gamma_1 \zeta^i_1 + \int_0^{\min(s_i,t)} \! \beta_1(t,t^\prime,s_i) z^i_1(t^\prime) \rmd t^\prime  \Bigg\}.
\end{align}

\subsubsection{Results.}
Figure~\ref{fig:PE-Liver} shows prediction error $\widehat{\mathrm{PE}}(u|t)$ as a function of $u$ for a fixed base time $t=3$ years for all four models. Again, each value of $\widehat{\mathrm{PE}}(u|t)$ is an average over the 20 random splits of the data into training and test data sets. The four models show similar prediction error up to $u=7$ years. After this point, retarded kernel model B has slightly lower prediction error than the other models. For example, at $t=9.2$ years, the joint model has $\widehat{\mathrm{PE}}(u|t)=0.236$, the landmark model has $\widehat{\mathrm{PE}}(u|t)=0.229$, model A has $\widehat{\mathrm{PE}}(u|t)=0.223$ and model B has $\widehat{\mathrm{PE}}(u|t)=0.208$.

Plots of average $\widehat{\mathrm{PE}}(u|t)$ against base time $t$ are shown in Figure~\ref{fig:PW-Liver} for fixed prediction windows $w_1=1$ year, $w_2=2$ years and $w_3=3$ years. For all three windows the four models exhibit very similar accuracy levels, with no model showing consistently superior predictions. 

For the liver data set the above results suggest that the retarded kernel models have a predictive accuracy that is comparable to those of standard methods.

\section{Summary and Discussion}
\label{Discuss}

In this work we propose a retarded kernel approach to dynamic prediction in survival analysis. In terms of complexity, our method comes somewhere between the two standard approaches, joint modelling and landmarking. It is more parsimonious than joint modelling, as it does not model the longitudinal covariate trajectory, and it makes no assumptions about the base hazard rate. This makes the method more practical for data with multiple time-dependent covariates. The retarded kernel approach conditions only on the observed covariates and, unlike joint modelling, makes no assumptions about covariate values in the future. This makes it more suitable for covariates that cannot easily be predicted, such as categorical ones. Compared to landmarking, the retarded kernel approach makes use of more of the available data. In landmarking, a new model is fitted at each landmark time, discarding individuals in the data set who have experienced the event before this landmark time. Additionally, landmarking only uses covariate measurements that are most recent before the landmark time. In contrast, the retarded kernel approach fits a single model that incorporates information from all individuals in the data set, using the full history of their covariate measurements. 

The retarded kernel approach relies on parameterisation of the association kernels $\beta_\mu(t,t^\prime\!,s)$. In this work we focused on two specific  parameterisations, motivated by  practical considerations. We required that our models reduce to the standard Cox model for static covariates, and that they  contain the instantaneous Cox model as a special case, so that they are natural extensions of familiar models.  However, alternative parameterisations or extensions (if demanded by the data at hand) can be incorporated without much effort. For example, one could include a `hard' time delay between covariate variations and their effect on hazard, or use parameterisations that favour time-translation invariance over consistency with standard Cox models. 

In tests on medical data, we found that the retarded kernel approach performs similarly to the two standard approaches in terms of predictive accuracy. Depending on the data set, base time and prediction window, each method (joint modelling, landmarking, or retarded kernels) had at some point the highest or the lowest prediction error; none appeared to be consistently superior or inferior across the scenarios we tested. These initial comparisons indicate that the retarded kernel method is a reasonable approach to dynamic prediction, worthy of further research and development.

There is scope to further develop the retarded kernel approach. For example, we used a na{\"i}ve interpolation procedure (step functions) but could  try  smoother interpolation methods such as Gaussian convolutions.\cite{Press:1992, VanDenBoomgaard:2001}  We could also take a Bayesian inference approach, using non-informative prior distributions or incorporating existing knowledge into informative ones. While joint modelling takes into account measurement errors, we have not attempted to do this for the retarded kernel model. Cox models were indeed found to be biased in the presence of such errors.\cite{Prentice:1982, Tsiatis:2001, Arisido:2019}  Hence, future work could involve building measurement error effects into the retarded kernel approach. One could also build on methods from WCE models\cite{Abrahamowicz:2012} and use a wider class of association kernels, for example those estimated via spline functions.  

In this work we compared against standard landmarking models, though extensions to these models exist.\cite{vanH:2011, Zhu:2020, Rizopoulos:2017} Similarly, we only considered joint models with instantaneous dependence on the `true' covariate trajectory $m^i_\mu(t)$ in the hazard function. This study  serves as a `proof of concept' and as a starting point for future investigations; we leave systematic comparison of alternative model variants to future work. Such comparative research could benefit from recent developments in simulation methods for dynamic predictions with time-varying covariates.\cite{Zhu:2018} Generating data according to the retarded kernel model with dependence on the period over which covariates are observed is non-trivial, but   could possibly be achieved by extending the permutational algorithm developed by Sylvestre and Abrahamowicz (2008).\cite{Sylvestre:2008} Such simulations could provide valuable tests for internal consistency. 

In summary, we have developed a `retarded kernel' approach to dynamic prediction that overcomes some limitations of existing methods. By conditioning the hazard rate on observed covariates over a given time frame, it  offers a simpler alternative to joint models without disregarding portions of longitudinal covariate data, as is the case with landmarking methods. Using three different clinical data sets we have demonstrated that retarded kernels can have a predictive accuracy comparable to that of established methods. We therefore believe that the retarded kernel method is a promising addition to the toolbox of dynamic prediction methods.

\begin{acks}
ACCC would like to thank Marianne Jonker for valuable discussions. 
\end{acks}

\subsection*{\normalsize\sagesf\bfseries Declaration of conflicting interests}\begin{refsize}\noindent 
The Authors declare that there is no conflict of interest.
\end{refsize}

\subsection*{\normalsize\sagesf\bfseries Funding}\begin{refsize}\noindent 
AD acknowledges funding by the Engineering and Physical Sciences Research Council (EPSRC UK), grant EP/R513131/1. TG is grateful for partial financial support by the Maria de Maeztu Program for Units of Excellence in R\&D (MDM-2017-0711).
\end{refsize}

\subsection*{\normalsize\sagesf\bfseries Supplementary material}\begin{refsize}\noindent 
For technical appendices and additional results the reader is referred to the Supplementary Materials at the GitHub repository \url{https://github.com/AnnieDavies/Supplement_Davies_Coolen_Galla_2021}. 
\end{refsize}

\subsection*{\normalsize\sagesf\bfseries Data availability statement}\begin{refsize}\noindent 
\texttt{C++} and \texttt{R} codes used to perform the data analysis in this manuscript are available at the GitHub repository \url{https://github.com/AnnieDavies/Supplement_Davies_Coolen_Galla_2021}. The data sets analysed are available publicly via the \texttt{JMbayes} \texttt{R} package.\cite{Rizo:2016}
\end{refsize}

%%BibTeX:
%%Harvard (name/date)
%\bibliographystyle{SageH}
%%Vancouver (numbered)
%\bibliographystyle{SageV}
%\bibliography{refs.bib}

%%BBL file:
%\input{main.bbl}

\end{document}